\documentclass[12pt]{article}
\advance \textheight by \topskip
\makeatletter
\@addtoreset{equation}{section}
 
\makeatother
\newcommand{\be}{\begin{equation}}
\newcommand{\ee}{\end{equation}}
\newcommand{\bea}{\begin{eqnarray}}
\newcommand{\eea}{\end{eqnarray}}
\newcommand{\dd}{\partial}

\def\>{\rangle}
\def\<{\langle}

\begin{document}

\title{
%\begin{flushright}
%{\small USACH-FM-01-02}\\[-0.4cm]
%{\small PM-01-07}\\[1cm]
%\end{flushright}
{\bf On the geometry of classically integrable two-dimensional non-linear sigma models}}

%Cubic root of translations in field theory }}

\author{
{\sf   N. Mohammedi} \thanks{e-mail:
nouri@lmpt.univ-tours.fr}$\,\,$${}$
%\thanks{This work was carried out at the Department of Applied Mathematics
%and Theoretical Physics, Cambridge, UK.} 
%{\sf  G. Moultaka }\thanks{e-mail:
%moultaka@lpm.univ-montp2.fr}$\,\,$${}^{b}$
% and
%{\sf M.~Rausch de Traubenberg}\thanks{e-mail:
%rausch@lpt1.u-strasbg.fr}$\,\,$$^{c}$\\
\\
%{\small {\it Universit\'e Fran\c{c}ois Rabelais de Tours,}}\\
{\small ${}${\it Laboratoire de Math\'ematiques et Physique Th\'eorique (CNRS - UMR 6083),}} \\
{\small {\it Universit\'e Fran\c{c}ois Rabelais de Tours,}}\\
{\small {\it Facult\'e des Sciences et Techniques,}}\\
{\small {\it Parc de Grandmont, F-37200 Tours, France.}}}
%\\ 
%{\small ${}^{b}${\it Laboratoire de Physique 
%Math\'ematique et Th\'eorique, CNRS UMR 5825, 
%Universit\'e Montpellier II,}}\\
%{\small {\it Place E. Bataillon, 34095 Montpellier,
%France}}\\
%{\small ${}^{c}${\it
%Laboratoire de Physique Th\'eorique, CNRS UMR  7085,
%Universit\'e Louis Pasteur}}\\
%{\small {\it  3 rue de
%l'Universit\'e, 67084 Strasbourg, France}}}
\date{}
\maketitle
\vskip-1.5cm

\vspace{2truecm}

\begin{abstract}

\noindent
A master equation expressing the zero curvature representation  of the equations
of motion of a two-dimensional
non-linear sigma models is found. The geometrical properties of this equation
are outlined. 
Special attention is paid to those representations possessing a spectral 
parameter.
Furthermore, a closer connection between integrability
and T-duality transformations is emphasised.
Finally, new integrable non-linear sigma models are found and all
their corresponding Lax pairs depend on a spectral parameter.

\end{abstract}

\newpage

%\section{The string effective action}
%\renewcommand{\theequation}{1.\arabic{equation}}   
%

\setcounter{equation}{0}

%\end{document}

\newpage

\section{Introduction}

The problem of finding dynamical systems which are 
integrable is a fascinating subject in mathematics and theoretical physics.
In classical mechanics integrability is understood as the possibility of finding
as many conserved quantities as the number of degrees of freedom of the dynamical
system. It happens, in some cases, that these conserved quantities lead
to the exact solvability of the associated equations of motion. 
In field theory, however, an infinite number of conserved charges is required
for integrability.

\par

The Lax formulation of integrability provides a method for constructing
conserved dynamical quantities. 
In this formulation, a two-dimensional field theory is considered to be classically integrable
if  a Lax pair $\left({\cal{A}}_0,{\cal{A}}_1\right)$ can be found such that the 
linear system{\footnote{Here, the two-dimensional coordinates are
$\left(\tau,\sigma\right)$ with $\dd_0={\dd\over \dd\tau}$ and $\dd_1={\dd\over \dd\sigma}$.
In the rest of the paper, however, we will use the complex coordinates
$\left(z=\tau +i\sigma\,,\,\bar z=\tau -i\sigma\right)$ together with
$\dd= {\dd\over\dd z}$ and $\bar\dd={\dd\over\dd\bar z}$.}}
\bea
\left[\dd_0+{\cal A}_0\left(\lambda\right)\right]\Psi &=&0
\nonumber\\
\left[\dd_1+{\cal A}_1\left(\lambda\right)\right]\Psi &=&0
\eea
yields, as its consistency condition,  the equations of motion of the 
two-dimensional theory under consideration. 
Here the matrices ${\cal{A}}_0$ and 
${\cal{A}}_1$ depend on the fields of the theory and possibly
on some free arbitrary parameter $\lambda$, known as the spectral parameter.
This parameter can be very useful in extracting conserved quantities.
The fields $\Psi$ can be either 
a column vector
or a matrix of the same dimension as ${\cal{A}}_0$ and ${\cal{A}}_1$.
The consistency condition (usually referred to as the
zero curvature condition) of this linear system is clearly
$\left\{\dd_0 {\cal A}_1 - \dd_1{\cal A}_0 
+\left[{\cal A}_0\,,\,{\cal A}_1\right]\right\}\Psi=0$.
\par
The conserved quantities are then constructed using the so-called 
monodromy matrix
\be
T\left(\lambda,\tau\right)= {\rm P}\exp\left(-\int_0^{2\pi}{\cal A}_1
\left(\lambda\,,\,\sigma\,,\,\tau\right)\,d\sigma
\right)\,\,\,\,,
\label{monodr}
\ee
where ${\rm P}$ stands for the path-ordered exponential and 
we have chosen $\sigma$ to be in the interval $[0\,,\,2\pi]$.
One can show that the traces of powers of the monodromy matrix,
${\rm Tr}\left[T^n\left(\lambda,\tau\right)\right]$, are independent
of the time $\tau$ and are in involution with respect to Poisson brackets:
$\left\{{\rm Tr}\left[T^m\left(\lambda_1,\tau\right)\right]\,,\,
{\rm Tr}\left[T^n\left(\lambda_2,\tau\right)\right]\right\}=0$. 
The proof of the first statement assumes the periodicity
condition ${\cal A}_0\left(\lambda\,,\,0\,,\,\tau\right)
= {\cal A}_0\left(\lambda\,,\,2\pi\,,\,\tau\right)$.
Expanding  ${\rm Tr}\left[T^n\left(\lambda,\tau\right)\right]$ in powers of $\lambda$ 
generates an infinite set of conserved charges (see \cite{faddeev,babelon} for more details).

\par
In this paper we would like to examine the question of integrability
in two-dimensional non-linear sigma models.  This is because there are only 
a handful cases of such theories which are known to be integrable 
(the principal chiral model, the Wess-Zumino-Witten model and their
various modifications \cite{balog1,evans,balog2,cherednik}).
It is therefore important to investigate whether other integrable 
models exist. Furthermore, the study of the properties of non-linear sigma models
involves often the geometry of the target space on which these theories are defined.
{}For instance, the renormalisation properties of these models constrains the geometry
of the target space \cite{friedan}. It will be shown in this paper that the requirement 
of integrability puts further constraints on the allowed target spaces.
This could be of crucial importance to string theory as non-linear 
sigma models are supposed to describe the propagation of the massless modes of bosonic
string theory \cite{gsw}. In other words, the conditions for conformal invariance at the quantum
level (the vanishing of the beta functions)
and the requirement of classical integrability of non-linear sigma models
might reduce the number of possibilities for the spaces on which one can carry 
out the compactification of the extra dimensions of string theory.      

\par
We start this paper by giving the general framework 
for a zero curvature representation of the equations of motion of
a two-dimensional non-linear sigma model. We derive a target space condition for 
this requirement  and analyse its resulting geometry.  In section 3, 
we provide some known solutions to this condition.
Section 4 deals with the issue of introducing a spectral parameter
in the Lax pair construction and further geometrical properties
are analysed there. We then study, in section 5, the integrability 
of a non-linear sigma model which generalises the principal
chiral sigma model. In section 6, the interplay between T-duality
and integrability of non-linear sigma models is explored.
This work is a continuation of an earlier
investigation \cite{moh}.

\section{Zero curvature representation of non-linear sigma models}

A two-dimensional non-linear sigma model is 
an interacting theory for some scalar fields 
$\varphi^i\left(z\,,\bar z\right)$ as
described
by the action 
\be
S=\int{\rm d}z{\rm d}\bar z \,Q_{ij}\left(\varphi\right)
\dd\varphi^i\bar\dd\varphi^j\,\,\,.
\label{sigma}
\ee
The metric and the anti-symmetric tensor fields of this theory
are defined as
\be
g_{ij}={1\over 2}\left(Q_{ij}+Q_{ji}\right)\,\,\,\,\,,\,\,\,\,\,
b_{ij}={1\over 2}\left(Q_{ij}-Q_{ji}\right)\,\,\,.
\ee
We will assume that the metric $g_{ij}$ is invertible 
and its inverse is denoted $g^{ij}$. Indices are raised and lowered
using this metric. We will also define, respectively, the Christoffel
symbols, the torsion and the generalised connection as follows  
\bea
\Gamma^k_{ij} &=& {1\over 2}g^{kl}
\left(\dd_ig_{lj}+\dd_jg_{li}-\dd_lg_{ij}\right)
\nonumber\\
H^k_{ij} &=& {1\over 2}g^{kl}
\left(\dd_lb_{ij}+\dd_jb_{li}+\dd_ib_{jl}\right)
\nonumber\\
\Omega^k_{ij} &=& \Gamma^k_{ij} - H^k_{ij}\,\,\,\,\,.
\eea
The equations of motion of this theory can be written as
\be
{\cal{E}}^l\,\,\equiv\,\,
\bar\dd\dd\varphi^l +\Omega^l_{ij}
\dd\varphi^i
\bar\dd\varphi^j=0\,\,\,\,.
\label{eqofmot}
\ee
%where $\Omega^i_{jk}$ is the generalised connection given by
%\be
%\Omega^i_{jk}=g^{il}\left(\dd_j Q_{lk}+\dd_kQ_{jl}-\dd_lQ_{jk}
%\right)\,\,\,\,
%\ee
%and $g^{ij}$ is the inverse of the metric tensor and is used to
%raise and lower indices.
Let us now construct a linear system whose 
consistency conditions are equivalent to these equations of motion
(a zero curvature representation). 
We take, as an ansatz,  this linear system to have the following 
form 
%$\Psi$
%\bea
%\left(\dd + {\cal{A}}\right)\Psi &=&0\nonumber\\
%\left(\bar\dd + \bar{\cal{A}}\right)\Psi &=&0\,\,\,\,,
%\label{lax}
%\eea
%where ${\cal{A}}$ and $\bar{\cal{A}}$ are two matrices whose form 
%is to be determined shortly. The fields $\Psi$ can be either 
%a column vector
%or a matrix of the same dimension as ${\cal{A}}$ and $\bar{\cal{A}}$.
%The consistency condition (usually reffered to as the
%zero curvature condition) of this linear system is clearly
%$\left\{\dd \bar {\cal A} - \bar\dd{\cal A} 
%+\left[{\cal A}\,,\,\bar{\cal A}\right]\right\}\Psi=0$.
%If this condition is to yield the equations motion
%(\ref{eqofmot}), then the two gauge fields ${\cal{A}}$ and $\bar{\cal{A}}$
%must have the form
\bea
\left[\dd + \alpha_i\left(\varphi\right)\dd\varphi^i\right]\Psi &=& 0
\nonumber\\
\left[\bar\dd + \beta_j\left(\varphi\right)\bar\dd\varphi^j\right]\Psi &=& 0
\,\,\,\,,
\label{AA}
\eea
where $\alpha_i$ and $\beta_i$ are two
matrices depending on the fields $\varphi^i$.
This form of the Lax pair is dictated by the fact that the equations of motions 
of the non-linear sigma model do not contain terms involving
$\dd^2$ or $\bar\dd^2$. 
\par
The compatibility condition of the linear system
takes then the form
\bea
{\cal{F}} \,\,\equiv \,\,
\left(\beta_i-\alpha_i\right)\bar\dd\dd\varphi^i
+\left(\dd_i\beta_j-\dd_j\alpha_i +\left[\alpha_i\,,\,\beta_j\right]\right)
\dd\varphi^i\bar\dd\varphi^j =0
\,\,\,\,\,.
\label{compat1}
\eea
The non-linear sigma model enjoys a zero curvature representation 
of its equations of motion 
if this compatibility condition can be written as
\be
{\cal{F}}\,\,= {\cal{E}}^i \mu_i  =0
\,\,\,\,
\label{compat2}
\ee
for some matrices $\mu_i\left(\varphi\right)$. In order
for this last relation to yield ${\cal{E}}^i=0$ as the only 
non trivial possibility, the
matrices $\mu_i$ have to be linearly independent and their
number must be equal to the dimension of the target space
of the non-linear sigma model.
\par
The compatibility condition of the linear system yields the
equations of motion of the two-dimensional non-linear sigma model,
that is equation (\ref{compat2}) holds,  
provided that the matrices $\alpha_i\left(\varphi\right)$,
$\beta_i\left(\varphi\right)$ and $\mu_i\left(\varphi\right)$
satisfy\footnote{These equations are found by comparing the terms involving 
$\dd\bar\dd\varphi^i$ and $\dd\varphi^i\bar\dd\varphi^j$ on both sides
of (\ref{compat2}).}
\bea
&& \beta_i-\alpha_i=
\mu_i\nonumber\\
&& \dd_i\beta_j -\dd_j\alpha_i +\left[\alpha_i\,,\beta_j\right]
=\Omega^l_{ij}\,\mu_l\,\,\,\,.
\label{set1}
\eea
%
%Let us now consider the following curvature
%\bea
%{\cal{A}}&\equiv &\left[\dd + \alpha_i\left(\varphi\right)\dd\varphi^i
%\,,\,
%\bar\dd + \beta_j\left(\varphi\right)\bar\dd\varphi^j
%\right]
%\nonumber\\
%&=&
%\left(\beta_i-\alpha_i\right)\bar\dd\dd\varphi^i
%+\left(\dd_i\beta_j-\dd_j\alpha_i +\left[\alpha_i\,,\,\beta_j\right]\right)
%\dd\varphi^i\bar\dd\varphi^j
%\,\,\,\,\,,
%\eea
%where $\alpha_i$ and $\beta_i$ are some field dependent
%matrices. 
%\par
%We would like to have a relation of the form
%\be
%{\cal{A}}= \mu_i{\cal{E}}^i\,\,\,\,
%\ee
%for some linearly independant matrices $\mu_i\left(\varphi\right)$. This is 
%the case 
%provided that the matrices $\alpha_i\left(\varphi\right)$,
%$\beta_i\left(\varphi\right)$ and $\mu_i\left(\varphi\right)$
%satisfy
%\bea
%&& \beta_i-\alpha_i=
%\mu_i\nonumber\\
%&& \dd_i\beta_j -\dd_j\alpha_i +\left[\alpha_i\,,\beta_j\right]
%=\Omega^l_{ij}\mu_l\,\,\,\,.
%\label{set1}
%\eea
%
%\par
%
%
%
%We will now return to our set of equations 
%in (\ref{set1}).
 
The first equation gives simply  $\beta_i$ in terms
of $\alpha_i$ and $\mu_i$ 
\be
\beta_i=\alpha_i+\mu_i\,\,\,\,. 
\ee
The second equation of the above set
can then be written as
\bea    
F_{ij} = -\left(\nabla_i\mu_j-\Omega^k_{ij}\,\mu_k\right)\,\,\,\,\,,
\label{F=Dmu}
\eea
where we have introduced, for later use,  the field strength $F_{ij}$ and the gauge 
covariant derivative corresponding to the matrices $\alpha_i$ 
\bea
F_{ij} &=& \dd_i\alpha_j -\dd_j\alpha_i+\left[\alpha_i\,,\,
\alpha_j\right]
\nonumber\\
\nabla_i X &=& \dd_i X+\left[\alpha_i\,,\, X\right]\,\,\,\,,
\eea
where $X$ denotes any matrix valued quantity. 
\par
Equation (\ref{F=Dmu}) is at the centre of the ability
to represent the equations of motion  
of a non-linear sigma model as a zero curvature 
condition of a linear system. The unknowns of the problem are
the two sets of matrices $\alpha_i$ and $\mu_i$ and the
generalised connection $\Omega^k_{ij}$. Each triplet
$\left(\alpha_i\,,\,\mu_i\,,\,\Omega^k_{ij}\right)$ satisfying
(\ref{F=Dmu}), yields a non-linear sigma model
with a zero curvature representation
(provided that one can extract $g_{ij}$ and $b_{ij}$
from the knowledge of $\Omega^k_{ij}$). However,
equation (\ref{F=Dmu}) does not guarantee that the
matrices $\alpha_i$ and $\mu_i$ will depend on a 
spectral parameter (which plays an important role
in the construction of the conserved quantities
of two-dimensional integrable theories).
Let us now explore some properties of this central equation.

\noindent
\vskip1.0cm
{\it{\Large{The geometry}}}
\vskip1.0cm

The consistency relation ($\dd_i\dd_j\mu_k-\dd_j\dd_i\mu_k=0$)
of equation
(\ref{F=Dmu}) is given by
\bea
{\cal{R}}^n_{\,\,\,\,jik}\,\mu_n &=& -{\cal D}_j F_{ik}
\nonumber\\
{\cal D}_j F_{ik} &\equiv& \nabla_jF_{ik}+\left[\mu_j\,,\,F_{ik}\right]
-\Omega^m_{ij}F_{mk}-\Omega^m_{kj}F_{im}\,\,\,\,,
\label{integrability}
\eea
where we have used the Bianchi identities
$\nabla_kF_{ij}+\nabla_jF_{ki}+ \nabla_iF_{jk}=0$.
Here ${\cal{R}}^n_{\,\,\,\,jik}$ is the generalised curvature tensor and
is defined by
\be
{\cal{R}}^n_{\,\,\,\,jik}=\dd_i\Omega^n_{kj}-\dd_k\Omega^n_{ij}
+\Omega^n_{im}\Omega^m_{kj}-\Omega^n_{km}\Omega^m_{ij}\,\,\,\,.
\ee
We notice immediately that if $F_{ij}=0$ (that is, 
$\alpha_i =M^{-1}\dd_i M$ for some invertible matrix 
$M(\varphi)$), then ${\cal{R}}^n_{\,\,\,\,jik}\,\mu_n=0$.
Since the matrices $\mu_i$ are assumed to be linearly independent,
we have ${\cal{R}}^n_{\,\,\,\,jik}=0$ and the target space of the non-linear
sigma model is, in this case,  parallelisable. 
\par
In the context of string theory, non-linear sigma models describe
the propagation of strings in non-trivial backgrounds. 
The consistency of this propagation is equivalent to the vanishing
of the beta functions of the non-linear sigma model \cite{callan, metsaev, shore}. 
At the one loop level, 
these beta functions (in the absence of the dilaton field) are characterised  
by the generalised Ricci tensor  ${\cal{R}}^n_{\,\,\,\,i}=g^{jk}{\cal{R}}^n_{\,\,\,\,jik}$. If 
the non-linear sigma model admits a zero curvature representation then
its generalised Ricci tensor satisfies
\bea
{\cal{R}}^n_{\,\,\,\,i}\,\mu_n &=& -{\cal D}^k F_{ik}\,\,\,\,\,.
\eea
In particular, if the generalised Ricci tensor vanishes
(namely, conformal invariance holds at the one loop level) then one has
\bea
{\cal D}^k F_{ik}=0\,\,\,\,.
\eea
This last equation is a generalisation of the equations of motion of pure 
non-Abelian Yang-Mills gauge theory.
\par
It is also interesting to split equation (\ref{F=Dmu}) into its symmetric  
and anti-symmetric parts. This yields
\bea
0 &=& \nabla_i\mu_j+\nabla_j\mu_i 
-2\Gamma^k_{ij}\,\mu_k
\nonumber\\
F_{ij} &=&  -{1\over 2}\left(\nabla_i\mu_j-\nabla_j\mu_i\right)
-H^k_{ij}\,\mu_k \,\,\,\,\,,
\eea
The first equation is a gauged version of a matrix valued Killing
equation. Indeed, if $\left[\alpha_i\,,\,\mu_j\right]+
\left[\alpha_j\,,\,\mu_i\right]=0$ then this first equation is 
simply $\dd_i\mu_j+\dd_j\mu_i 
-2\Gamma^k_{ij}\mu_k=0$. In this case the entries of
$\mu_i$ are Killing vectors (isometries) of the metric $g_{ij}$.
\par
The linear system (\ref{AA}) with $\beta_i=\alpha_i+\mu_i$ could be made
`more symmetric` by writing $\alpha_i=\gamma_i-{1\over 2}\mu_i$ and
$\beta_i=\gamma_i+{1\over 2}\mu_i$, for some matrices $\gamma_i(\varphi)$. 
The linear system takes then the form
\bea
\left[\dd + \left(\gamma_i-{1\over 2}\mu_i\right)\dd\varphi^i\right]\Psi &=& 0
\nonumber\\
\left[\bar\dd +\left(\gamma_j+{1\over 2}\mu_j\right) \bar\dd\varphi^j\right]\Psi &=& 0
\,\,\,\,.
\label{AA-sym}
\eea
In terms of the matrices $\gamma_i$, the symmetric and anti-symmetric parts
of equation (\ref{set1}) give
%\bea
%&&\left(\dd_i\gamma_j -\dd_j\gamma_i+\left[\gamma_i\,,\,\gamma_j\right]
%-{1\over 4}\left[\mu_i\,,\,\mu_j\right]\right)
%+{1\over 2}\left(\dd_i\mu_j+\dd_j\mu_i+\left[\gamma_i\,,\,\mu_j\right]
%+\left[\gamma_j\,,\,\mu_i\right]\right)
%\nonumber \\
%&&=\left(\Gamma^k_{ij} - H^k_{ij}\right)\,\mu_k\,
%\label{set1-sym}
%\eea
\bea
\dd_i\mu_j+\dd_j\mu_i+\left[\gamma_i\,,\,\mu_j\right]
+\left[\gamma_j\,,\,\mu_i\right] -2\Gamma^k_{ij}\,\mu_k &=&0
\nonumber \\
\dd_i\gamma_j -\dd_j\gamma_i+\left[\gamma_i\,,\,\gamma_j\right]
-{1\over 4}\left[\mu_i\,,\,\mu_j\right]
+H^k_{ij}\,\mu_k &=&0\,\,\,\,\,.
\label{set1-sym}
\eea
The advantage of working with the matrices $\gamma_i$ is that the derivatives
of $\mu_i$ do not appear in the anti-symmetric part of the last set of equations.

%\bea
%\dd_i\mu_j+\dd_j\mu_i+\left[\gamma_i\,,\,\mu_j\right]
%+\left[\gamma_j\,,\,\mu_i\right] -2\Gamma^k_{ij}\,\mu_k &=&0
%\nonumber \\
%\dd_i\gamma_j -\dd_j\gamma_i+\left[\gamma_i\,,\,\gamma_j\right]
%-{1\over 4}\left[\mu_i\,,\,\mu_j\right]
%+H^k_{ij}\,\mu_k &=&0\,\,\,\,\,.
%\eea
%The advantage of working with the matrices $\gamma_i$ is that the derivatives
%of $\mu_i$ do not appear in the anti-symmetric part of the last set of equations.

\par

Another geometric structure occurs when introducing the following change of variables
\bea
\mu_i &=& 2g_{ij}\,\nu^j=\left(Q_{ij}+Q_{ji}\right)\,\nu^j
\nonumber\\
\widetilde\alpha_i &=& \alpha_i+Q_{il}\,\nu^l\,\,\,\,\,\,.
\eea
In terms of the new variables $\nu^i$ and $\widetilde\alpha_i$, 
equation (\ref{F=Dmu}) takes the form
\bea
\nu^l\,\dd_lQ_{ij} + Q_{lj}\,\widetilde\nabla_i\nu^l 
+Q_{il}\,\widetilde\nabla_j\nu^l
=Q_{ik}Q_{lj}\left[\nu^k\,,\,\nu^l\right]
-\widetilde F_{ij}\,\,\,\,\,\,,
\label{K-S}
\eea
where $\widetilde F_{ij}=\dd_i\widetilde\alpha_j -\dd_j\widetilde\alpha_i
+\left[\widetilde\alpha_i\,,\,
\widetilde\alpha_j\right]$ and $\widetilde\nabla_j\nu^l=\dd_j\nu^l
+\left[\widetilde\alpha_j\,,\,\nu^l\right]$.
Notice that the left-hand side of this last equation is a gauged version
of a matrix valued Lie derivative for the tensor $Q_{ij}$.
With the new variables, $\widetilde\alpha_i$ and $\nu^i$,  the linear system is 
given by
\bea
\left[\dd +\left(-Q_{il}\,\nu^l+\widetilde\alpha_i\right)\dd\varphi^i\right]\Psi &=& 0
\nonumber\\
\left[\bar \dd +\left(Q_{kj}\,\nu^k+\widetilde\alpha_j\right)\bar\dd\varphi^j\right]\Psi &=& 0
\label{newlax}
\eea
Equation (\ref{K-S}), {\it {when ${\widetilde{\alpha}}_i=0$ and when $\nu^i$ take values in a Lie algebra}}, 
is precisely the relation encountered 
in the context of Poisson-Lie duality and whose solution was given by Klim\v{c}\'{\i}k and \v{S}evera
in \cite{klimcik1,klimcik2}. 
\par
Furthermore, equation (\ref{K-S}) can be interpreted in the following way: 
Let us choose two currents such that
\bea
I^a=\left(v^{-1}\right)^a_i\,\dd\varphi^i
\nonumber\\
\bar I^a=\left(w^{-1}\right)^a_i\,\bar\dd\varphi^i
\label{I-I}
\eea
with $v^i_a$ and $w^i_a$ two field-dependent matrices
and whose inverses are, respectively, 
$(v^{-1})^a_i$ and $(w^{-1})^a_i$. In terms of these
currents, the equations of motion of the non-linear sigma model (\ref{sigma})
%\be
%S=\int{\rm d}z{\rm d}\bar z \,Q_{ij}\left(\varphi\right)
%\dd\varphi^i\bar\dd\varphi^j\,\,\,
%\ee
are expressed as
\bea
{\cal E}_l\equiv Q_{lj}w^j_a\dd\bar I^a + Q_{il}v^i_a\bar\dd I^a
-\left[v^i_aw^j_b\dd_lQ_{ij} -v^k_a\dd_k\left(Q_{lj}w^j_b\right)
-w^k_b\dd_k\left(Q_{il}v^i_a\right)\right]I^a\bar I^b=0\,\,\,\,.
\eea
In addition, these currents satisfy the Bianchi identities (stemming from the identity
$\dd\bar\dd\varphi^i - \bar\dd\dd\varphi^i=0$ in (\ref{I-I}))
\bea
{\cal B}^l\equiv w^l_a\dd\bar I^a -  v^l_a\bar\dd I^a
+\left(v^k_a\dd_kw^l_b - w^k_b\dd_kv^l_a\right)I^a\bar I^b=0\,\,\,\,.
\eea
In terms of these currents the Lax pair (\ref{AA})
%\bea
%\left[\dd + \alpha_i\left(\varphi\right)\dd\varphi^i\right]\Psi &=& 0
%\nonumber\\
%\left[\bar\dd + \beta_j\left(\varphi\right)\bar\dd\varphi^j\right]\Psi &=& 0
%\,\,\,\,,
%\eea
takes the form
%\bea
%\left[\dd +\left(-Q_{il}\,\nu^l+\widetilde\alpha_i\right)v^i_a I^a\right]\Psi &=& 0
%\nonumber\\
%\left[\bar \dd +\left(Q_{kj}\,\nu^k+\widetilde\alpha_j\right) w^j_b \bar I^b\right]\Psi &=& 0
%\eea
\bea
\left[\dd + \alpha_i v^i_a I^a\right]\Psi &=& 0
\nonumber\\
\left[\bar\dd + \beta_j w^j_b \bar I^b \right]\Psi &=& 0
\,\,\,\,.
\eea 
The consistency condition (the zero curvature condition) of this Lax pair is
\bea
{\cal Z} \equiv w^j_b \beta_j \dd\bar I^b -  v^i_a \alpha_i \bar\dd I^a
+\left\{v^k_a\dd_k\left(w^j_b\beta_j\right)
- w^k_b\dd_k\left(v^i_a\alpha_i\right) + v^i_aw^j_b
\left[\alpha_i\,,\,\beta_j\right]\right\}I^a\bar I^b=0\,\,\,\,.
\eea
Equation (\ref{K-S}) amounts then to demanding that
\bea
{\cal Z}=\nu^l{\cal E}_l + \widetilde{\alpha}_l{\cal B}^l\,\,\,\,\,,
\label{E-B}
\eea
where $\alpha_i = -Q_{il}\nu^l + \widetilde{\alpha}_i$ and  
and $\beta_i = Q_{li}\nu^l + \widetilde{\alpha}_i$.
This means that the Lax pair yields the equations of motion of the non-linear 
sigma model up to terms which identically vanish.
The importance of working with currents will show up in the rest of the paper.

%It is clear that the integrability equation (\ref{F=Dmu}) leads to some
%interesting geometrical structures and deserves further studies.

%\par
%Finally, equations (\ref{set1-sym}) for $\mu_i=2g_{ij}\,\nu^j$ transform into
%\bea
%\nu^l\dd_l\,g_{ij} + g_{lj}\,\dd_i\nu^l + g_{il}\,\dd_j\nu^l
%+g_{il}\left[\gamma_j\,,\,\nu^l\right] 
%+g_{lj}\left[\gamma_i\,,\,\nu^l\right] 
%&=&0
%\nonumber \\
%\dd_i\gamma_j -\dd_j\gamma_i+\left[\gamma_i\,,\,\gamma_j\right]
%-g_{ik}g_{jl}\left[\nu^k\,,\,\nu^l\right]
%+2\,H_{ijk}\,\nu^k &=&0\,\,\,\,\,.
%\eea

\section{Known Solutions}

As stated above, all the quantities entering equation
(\ref{F=Dmu}) are unknowns. In order to find some 
solutions, we proceed by fixing some of these unknowns.

As a start, let us first check that this formalism reproduces
the two well-known integrable non-linear sigma models, namely the
principle chiral model and the Wess-Zumino-Witten model.
These models are found by taking the following expressions
for the matrices $\alpha_i$ and $\mu_i$
\bea
\alpha_i=x\,g^{-1}\dd_ig\,\,\,\,\,,\,\,\,\,\,\,
\mu_i=y\,g^{-1}\dd_ig\,\,\,\,,
\eea
where $g\left(\varphi\right)$ is a Lie group element corresponding
to some Lie algebra ${\cal{G}}$ defined by the commutation
relations $\left[T_a\,,\,T_b\right]=f^c_{ab}T_c$.
The indices of the Lie algebra $a\,,b\,,c\,,\dots$ have the same 
range as those of the target space of the sigma model $i\,,j\,,k\,,\dots$.
We will use the fact that the gauge connection 
$A_i=g^{-1}\dd_ig=e^a_i\left(\varphi\right)T_a$
satisfies the Bianchi identity $\dd_i A_j -\dd_j A_i+\left[A_i\,,\,
A_j\right]=0$. The inverses of the vielbiens $e^a_i$ are denoted
$E^i_a$ and satisfy $e^a_iE^i_b=\delta^a_b$ and $e^a_iE^j_a=\delta^j_i$.
Finally, the quantities $x$ and $y$ are two constant parameters which will 
provide the spectral parameter. We assume that $x$ and $y$ are different from zero.
\par
Injecting the expressions of $\alpha_i$ and $\mu_i$
in (\ref{F=Dmu}) leads to
\bea
\Gamma^k_{ij} &=& {1\over 2}E^k_a\left(\dd_ie^a_j+ \dd_je^a_i\right)
\nonumber\\
H^k_{ij} &=& 
%-{1\over y}\left(x^2-x+xy-{1\over 2}y\right)
\kappa\,\, e^a_ie^b_jE^k_c\,f^c_{ab}\,\,\,\,,
\label{gam-h}
\eea
where $\kappa=-{1\over y}\left(x^2-x+xy-{1\over 2}y\right)$.
%
%In order for the torsion $H^k_{ij}$ to be independent of the parameters
%$x$ and $y$ (the spectral parameters), we require that
%$-{1\over y}\left(x^2-x+xy-{1\over 2}y\right)=\kappa$ for some parameter
%$\kappa$. This allows us to express $y$ in terms of $x$ as
%$y=\left(x^2-x\right)/\left({1\over 2}-x-\kappa\right)$. The parameter
%$x$ is then the spectral parameter. If we did not demand the independence

\par

The above Christoffel symbols $\Gamma^k_{ij}$ and torsion $H^k_{ij}$
are those corresponding to the following metric $g_{ij}$ and anti-symmetric 
tensor $b_{ij}$
\bea
g_{ij} &=& \eta_{ab}e^a_ie^b_j
\nonumber\\
H_{ijk} &=& \kappa\,\,\eta_{da}f^d_{bc}e^b_ie^c_je^a_k
\,\,\,\,\,,
\label{wzw1}
\eea
where $\eta_{ab}$ is an invertible bilinear form of the Lie
algebra $\cal{G}$ satisfying $\eta_{ab}f^b_{cd}+\eta_{cb}f^b_{ad}=0$. 
%The torsion $H_{ijk}$ is then
%given by
%\be
%H_{ijk}=\kappa\,\,\eta_{da}f^d_{bc}e^b_ie^c_je^a_k\,\,\,\,\,.
%\label{wzw2}
%\ee
Owing to the property that $\dd_ie^a_j -\dd_je^a_i
+f^a_{bc}e^b_ie^c_j=0$, the torsion is a closed 
three form. Therefore $b_{ij}$ exists locally.
\par
To summarise, the Lax pair construction
for the class of theories represented by the metric and the torsion in (\ref{wzw1}) 
is given by  
\bea
\left[\dd + x\,\left(g^{-1}\dd_i g\right)\,\dd\varphi^i\right]\Psi &=&0
\nonumber\\
\left[\bar\dd + {x(2\kappa +1)\over 2x+2\kappa -1}
\,\left(g^{-1}\dd_j g\right)\,\bar\dd\varphi^j\right]\Psi &=&0\,\,\,\,,
\label{lax1}
\eea
with $x$ being the spectral parameter
and $\kappa$ a parameter defining the different models.
The class of non-linear sigma models defined by (\ref{wzw1})
includes the principal chiral sigma model 
($\kappa=0$); 
the Wess-Zumino-Witten model ($\kappa={1\over 2}$);
and the non-conformally invariant  
Wess-Zumino-Witten model ($\kappa\ne{1\over 2}$). 
%By writing $x=1/(1+\lambda)$ one gets the most common notation for the
%Lax representation of these models
%\bea
%\left[\dd + {1\over 1+\lambda}\,\left(g^{-1}\dd_i g\right)\,\dd\varphi^i\right]\Psi &=&0
%\nonumber\\
%\left[\bar\dd + {2\kappa +1\over (1+\lambda)(2\kappa -1) +2}
%\,\left(g^{-1}\dd_j g\right)\,\bar\dd\varphi^j\right]\Psi &=&0\,\,\,\,,
%\label{lax1}
%\eea
\par
Another interesting theory is found  when the matrices $\alpha_i$ and 
$\mu_i$ are constant. In this case we take 
\bea
\alpha_i=x\,T_i\,\,\,\,\,\,,\,\,\,\,\,\mu_i=y\,T_i
\,\,\,\,\,,
\eea 
where $\left[T_i\,,\,T_j\right]=f_{ij}^k\,T_k$. Replacing these in 
equation (\ref{F=Dmu}) yields 
\bea
\Gamma^k_{ij} &=& 0\nonumber\\
\,H^k_{ij} &=& \rho f^k_{ij}\,\,\,\,,
\eea
where $\rho=-{1\over y}\left(x^2+xy\right)$.
\par
These relations yield a non-linear sigma model defined by
\bea
g_{ij} &=& \eta_{ij}
\nonumber \\
b_{ij} &=& {2\over 3}\,\rho\,\eta_{kl}f^l_{ij}\,\varphi^k
\,\,\,\,\,,
\label{nappi}
\eea
where $\eta_{ij}$ 
is the invertible bilinear form 
corresponding to the Lie algebra $\left[T_i\,,\,T_j\right]=f_{ij}^k\,T_k$
($\eta_{ij}$ must satisfy $\eta_{kl}f^l_{ij}+\eta_{il}f^l_{kj}=0$ 
in order
for $H_{ijk} = \rho\,\eta_{kl}f^l_{ij}$ to be totally anti-symmetric).
\par
Therefore, the linear system for this non-linear sigma model (\ref{nappi})
is given by
\bea
\left[\dd + x\,T_i\,\dd\varphi^i\right]\Psi &=&0
\nonumber\\
\left[\bar\dd +{\rho x\over x+\rho}\,T_j\,\bar\dd\varphi^j\right]\Psi &=&0\,\,\,\,,
\label{lax2}
\eea
where $x$ plays the role of the spectral parameter while $\rho$
is a free parameter\footnote{In fact the parameter $\rho$ has 
no physical meaning as it can be simply absorbed by a 
rescaling of the field $\varphi^i$. This operation leads to an overall factor 
in the Lagrangian.}. 
The quantum properties
of this model, thought for a while to be the dual of the
principal chiral sigma model \cite{nadual1,nadual2},  have been studied in \cite{nappi}.
\par
In this paper, we will present other non-linear sigma models
which admit a Lax pair representation. Some of these models are
new integrable two-dimensional theories.

\section{Construction having a multiplicative spectral parameter}

So far we have only given the conditions under which the equations of motion of
a non-linear sigma model admit a zero curvature representation. 
However, as mentioned earlier these conditions (namely, equation (\ref{F=Dmu}))
do not guarantee the existence of a spectral parameter. {}For instance, the
non-linear sigma model constructed in the context of Poisson-Lie 
T-duality \cite{klimcik1,klimcik2} admits a zero curvature representation 
but without a spectral parameter. Therefore, the class of non-linear sigma
models enjoying a zero curvature representation is necessarily larger than
the class of non-linear sigma models possessing the same property but 
with a spectral parameter. 
\par
As the presence of a spectral parameter in the Lax pair is of crucial
importance in extracting conserved quantities \cite{faddeev,babelon}, we will now demand that
our construction depends on such a spectral parameter. 
\par
We assume, in this section, that
this spectral parameter enters in a multiplicative manner. Namely, we 
consider the following 
ansatz for the linear system 
\bea
\left[\dd + x\,\widehat\alpha_i\left(\varphi\right)\dd\varphi^i\right]\Psi &=& 0
\nonumber\\
\left[\bar\dd + y(x)\,\widehat\beta_j\left(\varphi\right)\bar\dd\varphi^j\right]\Psi &=& 0
\,\,\,\,,
\label{multiplicative}
\eea
where $x$ is our arbitrary spectral parameter and $y(x)$ is a function of $x$.
The matrices $\widehat\alpha_i$ and $\widehat\beta_i$ are independent of $x$. 
\par
The equivalence of the compatibility condition of the linear system and 
the equations of motion of the non-linear sigma model 
(${\cal F}=\mu_l{\cal E}^l$)
results in the 
equality
\bea 
\left(y\widehat\beta_i-x\widehat\alpha_i\right)\bar\dd\dd\varphi^i
+\left(y\dd_i\widehat\beta_j-x\dd_j\widehat\alpha_i 
+xy\left[\widehat\alpha_i\,,\,\widehat\beta_j\right]\right)
\dd\varphi^i\bar\dd\varphi^j 
=
\mu_l\left(
\bar\dd\dd\varphi^l +\Omega^l_{ij}
\dd\varphi^i
\bar\dd\varphi^j\right)\,\,\,\,.
\eea
Upon identification of the terms involving $\bar\dd\dd\varphi^i$
and $\dd\varphi^i\bar\dd\varphi^j$ on both sides of this equation we get
\bea
&& \mu_i=y\widehat\beta_i-x\widehat\alpha_i
\nonumber\\
&&y \dd_i\widehat\beta_j -x\dd_j\widehat\alpha_i 
+xy\left[\widehat\alpha_i\,,\widehat\beta_j\right]
=\Omega^l_{ij}\,\widehat\mu_l\,\,\,\,.
\eea
The first equation of this set determines the matrix $\mu_i$
while the second leads to
\bea
y\left(\dd_i\widehat\beta_j -\Omega^l_{ij}\widehat\beta_l\right)
-x\left(\dd_j\widehat\alpha_i - \Omega^l_{ij}\widehat\alpha_l\right)
+xy\left[\widehat\alpha_i\,,\widehat\beta_j\right]
=0\,\,\,\,.
\label{ABC}
\eea
The important point here is that this last equation should hold for {\it any} 
value of the spectral parameter $x$
(we recall that $\widehat\alpha_i$, $\widehat\beta_i$ and $\Omega^l_{ij}$ are independent
of $x$). 
This requirement is fulfilled only if\footnote{
Equation (\ref{ABC}) is of the form $yA-xB+xyC=0$ for three
$x$-independent matrices $A$, $B$ and $C$. There are, of course, various
cases  to be studied. We have analysed here only the case when
the three matrices are proportional to each other ($A=aC$, $B=bC$).
The other cases lead, in general, to known integrable 
non-linear sigma models. 
In the rest of the paper, we will assume that both $a$ and $b$ are different from zero.}
%We are assuming that 
%$\left[\widehat\alpha_i\,,\widehat\beta_j\right]\ne 0$.}}
\bea
\left(\dd_i\widehat\beta_j -\Omega^l_{ij}\widehat\beta_l\right) &=& 
a\left[\widehat\alpha_i\,,\widehat\beta_j\right]
\nonumber \\
\left(\dd_j\widehat\alpha_i - \Omega^l_{ij}\widehat\alpha_l\right)
&=& b\left[\widehat\alpha_i\,,\widehat\beta_j\right]
\,\,\,\,\,
\label{a-b}
\eea
together with
\be
y(x)={bx\over x + a}\,\,\,\,
\ee
for two parameters $a$ and $b$ (which will define the different
non-linear sigma models).
It is easy to see that by taking $\widehat\alpha_i =\widehat\beta_i=g^{-1}\dd_ig$,
$a=(2\kappa -1)/2$ and $b=(2\kappa +1)/2$ one recovers
the construction already given in (\ref{lax1}). Similarly, by taking $\widehat\alpha_i =\widehat\beta_i=T_i$
(where $\left[T_i\,,\,T_j\right]=f_{ij}^k\,T_k$)
and $a=b=\rho$ we arrive at the linear system in (\ref{lax2}). 
%In the rest of the paper, we will assume that both $a$ and $b$ are different from zero.
%The case $a=0$ leads to $y(x)=1$ and the integrable non-linear sigma model is the conformal
%Wess-Zumino-Witten model (that is, equation (\ref{lax1}) with $\kappa=1/2$) for which
%$\widehat\alpha_i =\widehat\beta_i=g^{-1}\dd_ig$ and $b=1$.

\noindent
\vskip1.0cm
{\it{\Large{The geometry}}}
\vskip1.0cm 

\noindent
It is instructive to point out the special features of the
theories which admit the Lax construction (\ref{multiplicative}). 
We would like, in particular, to provide an interpretation for
the two matrices $\widehat\alpha_i$ and $\widehat\beta_i$ of 
this Lax pair.
In order to do this, 
we start by combining the two equations in (\ref{a-b}) and taking the symmetric and anti-symmetric parts.
This yields
\bea
&&\dd_i K_j +\dd_j K_i -2\Gamma^l_{ij}\,K_l=0
\nonumber
\\
&&\dd_i L_j -\dd_j L_i + 2H^l_{ij}\,K_l = 0 
\nonumber 
\\
&&\dd_i L_j +\dd_j L_i -2\Gamma^l_{ij}\,L_l=
 \left[L_i\,,\,K_j\right]+\left[L_j\,,\,K_i\right]
\nonumber
\\
&& \dd_i K_j -\dd_j K_i + 2H^l_{ij}\,L_l
= 
\left[L_i\,,\,L_j\right]-\left[K_i\,,\,K_j\right]
\,\,\,\,\,,
\label{combination} 
\eea
where
\bea
L_i &=& b\,\widehat\beta_i+a\,\widehat\alpha_i
\nonumber\\
K_i &=& b\,\widehat\beta_i-a\,\widehat\alpha_i
\,\,\,\,\,.
\eea
%Recalling now that the torsion tensor, as defined in (),  is
%required to satisfy
%\bea
%\dd_lH_{ijk} -\dd_kH_{lij}
%+\dd_jH_{kli}-\dd_iH_{jkl}=0\,\,\,\,.
%\eea
The first two equations of the set (\ref{combination}) can be cast in the form
\bea
&& K^l\dd_lg_{ij} + g_{lj}\dd_iK^l + 
g_{il}\dd_j K^l=0
\nonumber\\
&& K^l\dd_lb_{ij} + b_{lj}\dd_iK^l + 
b_{il}\dd_j K^l=-\left[\dd_i\left(L_j+b_{jl}K^l\right)
- \dd_j\left(L_i+b_{il}K^l\right)\right]
%&& K^l\dd_lH_{ijk} + H_{ljk}\dd_iK^l + 
%H_{ilk}\dd_j K^l+ H_{ijl}\dd_kK^l=0
\,\,\,\,\,,
\eea
where $K^i=g^{ij}K_j$. In terms of the tensor $Q_{ij}=g_{ij}+b_{ij}$,  these
two equations combine to yield 
\bea
K^l\dd_lQ_{ij} + Q_{lj}\dd_iK^l + 
Q_{il}\dd_j K^l=-\left[\dd_i\left(L_j+b_{jl}K^l\right)
- \dd_j\left(L_i+b_{il}K^l\right)\right]
\,\,\,\,\,.
\eea
This last relation is precisely the condition needed for the
non-linear sigma model (\ref{sigma}) to enjoy the isometry symmetry
\cite{jack,hull}
\be
\delta\varphi^i=\varepsilon^{AB} K^i_{AB}\,\,\,\,\,,
\label{isometry}
\ee
where $K^i_{AB}$ are the entries of the matrix $K^i$ and $\varepsilon^{AB}$
are constant infinitesimal parameters. 
We conclude that if a non-linear sigma model accepts a Lax pair representation
of the form (\ref{multiplicative}) then this model possesses
an isometry symmetry (of course, for  $K^i\ne 0$).
\par
The conserved currents corresponding to the isometry transformation 
(\ref{isometry}) are
\bea
J &=& \left(K_i -L_i\right)\dd\varphi^i=
-{2a}\,\widehat{\alpha}_i\dd\varphi^i
\nonumber \\
\bar J &=& \left(K_i + L_i\right)\bar\dd \varphi^i=
{2b}\,\widehat{\beta}_i\bar\dd\varphi^i
\,\,\,\,.
\eea
In terms of the two currents $J$ and $\bar J$, the Lax pair (\ref{multiplicative})
is given by
\bea
\left[\dd - {x\over 2a}\, J\right]\Psi &=& 0
\nonumber\\
\left[\bar\dd + {x\over 2\left(x+a\right)}\,\bar J\right]\Psi &=& 0
\,\,\,\,.
\label{J-J}
\eea
Moreover, by contacting both sides of equations (\ref{a-b}) with $\dd\varphi^i\bar\dd\varphi^j$,
we obtain
\bea
\dd\bar J + {1\over 2}\left[J\,,\,\bar J\right] 
&=&2b\,\widehat{\beta}_l\,{\cal E}^l 
\nonumber\\
\bar \dd J - {1\over 2}\left[J\,,\,\bar J\right] 
&=&-2a\,\widehat{\alpha}_l\,{\cal E}^l \,\,\,\,, 
\eea
where ${\cal{E}}^l\,\,\equiv\,\,
\bar\dd\dd\varphi^l +\Omega^l_{ij}
\dd\varphi^i
\bar\dd\varphi^j=0$ are the equations of motion
of the non-linear sigma model. 
These relations can be cast into the form
\bea
\dd\bar J + \bar \dd J &=& 2K_i\,{\cal E}^i
\nonumber \\
\dd\bar J - \bar \dd J + 
\left[J\,,\,\bar J\right] &=& 2L_i\,{\cal E}^i
\,\,\,\,. 
\eea
This last set of equations, suggests the study of three different cases:
\par
\noindent
{\bf 1) $K_i=0$ and $L_i\ne 0$ :}
\newline 
In this case the two currents $J$ and $\bar J$ 
satisfy the equation $\dd\bar J + \bar \dd J=0$ independently of the 
equations of motion of the non-linear sigma model. Therefore, the
equation $\dd\bar J + \bar \dd J=0$ is a Bianchi identity (a topological
property) for the 
two currents.
Furthermore, equations (\ref{combination})
reduce to 
\bea
&&\dd_i L_j -\dd_j L_i  = 0 
\nonumber 
\\
&&\dd_i L_j +\dd_j L_i -2\Gamma^l_{ij}\,L_l=
 0
\nonumber
\\
&& 2H^l_{ij}\,L_l
= 
\left[L_i\,,\,L_j\right]
\,\,\,\,\,.
\eea
This set has a unique solution given by $L_i=2aT_i$, where 
$\left[T_i\,,\,T_j\right]=f^k_{ij}T_k$ and the integrable 
non-linear sigma model is the model defined in (\ref{nappi})
with the identification $a=\rho$. 
\par
\noindent
{\bf 2) $L_i=0$ and $K_i\ne 0$ :}
\newline 
Here the two currents $J$ and $\bar J$ are the conserved currents
corresponding to the isometry symmetry 
$\delta\varphi^i=\varepsilon^{AB}K^i_{AB}$
of the non-linear sigma
model and satisfy the Bianchi
identity 
$\dd\bar J - \bar \dd J + \left[J\,,\,\bar J\right]=0$. In this case
the set (\ref{combination}) gives
\bea
&&\dd_i K_j +\dd_j K_i -2\Gamma^l_{ij}\,K_l=0
\nonumber
\\
&& 2H^l_{ij}\,K_l = 0 
\nonumber 
\\
&& \dd_i K_j -\dd_j K_i + 
\left[K_i\,,\,K_j\right]=0
\,\,\,\,\,.
\eea
This has a unique solution given by $K_i=g^{-1}\dd_ig$,
for some Lie group element $g(\varphi)$,  and the integrable theory
is the principal chiral non-linear sigma model (that is, equation (\ref{wzw1})
with $\kappa=0$). The parameter $a$ in (\ref{J-J}) is equal to $-1/2$.
\par
\noindent
{\bf 3) $K_i\ne 0$ and $L_i\ne 0$ :}
\newline
This case means that  
the integrable non-linear sigma model must possess the isometry
symmetry (\ref{isometry}) whose conserved currents are $J$ and $\bar J$
(namely, $\dd\bar J + \bar \dd J=0$ on shell). 
Moreover, the field strength $\dd\bar J - \bar \dd J + \left[J\,,\,\bar J\right]$
vanishes only when the equations of motion are obeyed and is no longer a Bianchi
identity as in the previous case.
\par 
As seen earlier, the first two cases are unique and lead to known integrable non-linear
sigma models. Therefore, any new integrable non-linear sigma model 
(having (\ref{J-J}) as a Lax pair) must fit in this third class.
Notice that the conformal and non conformal WZW models are part of this third class.
Indeed,  by taking
$K_i= g^{-1}\dd_ig$ and 
$L_i=2\kappa \, g^{-1}\dd_ig$ (for $\kappa\ne 0$) and injecting them into
the set (\ref{combination}) one gets precisely equations (\ref{gam-h})
leading to the WZW models (\ref{wzw1}).
\par
It is therefore natural to ask whether other
non-linear sigma models, possessing (\ref{J-J}) as a Lax pair, exist. We will provide 
below an example of a non-linear sigma model, different from (\ref{wzw1}) and (\ref{nappi}), 
which have such a property.

%In other words, the equations of motion of the non-linear sigma model 
%guarantee that $\dd \bar J + \bar\dd J =0$. 
%\par
%Expressing $\widehat{\alpha}_i\dd\varphi^i$ and  $\widehat{\beta}_i\bar\dd\varphi^i$
%in terms of the two currents $J$ and $\bar J$, the zero curvature corresponding
%to the linear system () gives
%\bea
%\left[\dd -{x\over 2a}\,J\,,\,\bar\dd +{x\over 2\left(x+a\right)}\,\bar J\right]
%&=& 
%{x\left(x+2a\right)\over 4a\left(x+a\right)}\left(\dd \bar J + \bar\dd J\right)
%\nonumber \\
%&-&
%{x^2\over 4a\left(x+a\right)}\left(\dd \bar J - \bar\dd J+\left[J\,,\,\bar J\right]
%\right)
%=0
%\eea

\noindent
\vskip1.0cm
\noindent
{\it{\Large{An Example}}}
\vskip1.0cm

The non-linear sigma model we consider here is based on the $SU(2)$ Lie algebra.
Its action is given by 
\be
S =
\int{\rm d}z{\rm d}\bar z \,
\left\{\dd r\bar \dd r + B(r)\delta_{ab}\dd n^a\bar\dd n^b
+ C(r)\epsilon_{abc} n^a\dd n^b\bar \dd n^c \right\}
%+ 
%\chi\left(\delta_{ab}n^an^b -1\right)\right\} 
\,\,\,\,,
\ee
where $r$ and $n^a$ $(a=1,2,3)$ are the fields of the non-linear
sigma model subject to the constraint $\delta_{ab}n^an^b=1$. 
The $SU(2)$ Lie algebra is $\left[T_a\,,\,T_b\right]= \epsilon_{abc}T_c$
with
$\epsilon_{123}=1$. We shall make no distinction between upper and lower 
$SU(2)$ indices. 
\par
It is important to mention that this non-linear
sigma model was first studied in \cite{balog1} 
and later reexamined in \cite{evans}. The authors of 
\cite{balog1} demanded that this model should be: firstly classically integrable
and secondly it admits two commuting Kac-Moody algebras at the level of Poisson
brackets (the Hamiltonian is quadratic in the Kac-Moody currents). 
Here we will require that the model is only classically integrable.
Despite the fact that our requirement is less restrictive, the only solution
we found is that of \cite{balog1}.
\par
The equations of motion of this non-linear sigma model are
\bea
0&=&-2  \dd\bar\dd r  
+ B^\prime \delta_{ab} \dd n^a\bar\dd n^b + C^\prime \epsilon_{abc} n^a \dd n^b \bar \dd n^c 
\,\,\,\,,
\nonumber \\
0&=&-2B \dd\bar\dd n^b -B^\prime \left(\dd r\bar\dd n^b + \bar \dd r \dd n^b\right)
- 2 B n^b \delta_{cd}\dd n^c \bar\dd n^d  
\nonumber \\
&+& \epsilon_{bac}\left[3 C \dd n^a \bar \dd n^c + C^\prime n^a\left(\dd r\bar\dd n^c - \bar \dd r \dd n^c\right)
\right] -3 C \epsilon_{cde} n^b n^ c \dd n^d \bar \dd n^e \,\,\,\,.
\label{eom-su(2)}
\eea
%
%
%{\delta {\cal S}\over \delta n^b} &=&0=
%\delta_{ba}\left[
%-2B \dd\bar\dd n^a -B^\prime \left(\dd r\bar\dd n^a + \bar \dd r \dd n^a\right)
%+ 2 \chi n^a\right] 
%\nonumber \\
%&+& \epsilon_{bac}\left[3 \dd n^a \bar \dd n^c + C^\prime n^a\left(\dd r\bar\dd n^c - \bar \dd r \dd n^c\right)
%\right]
%\nonumber \\
%{\delta {\cal S}\over \delta \chi} &=&0 =\delta_{ab}n^an^b -1  \,\,\,\,.
%\eea
Here a prime denotes the derivative with respect to $r$. The equations of motion for the field $n^b$ are
obtained by introducing a Lagrange multiplier 
for the constraint $\delta_{ab}n^an^b=1$.
%Contracting the second equation with $n^b$ and using $\delta_{ab}n^an^b =1$, we get
%\be
%\chi = - B\delta_{ab}\dd n^a\bar \dd n^b -{3\over 2}\epsilon_{bac} n^b \dd n^a \bar \dd n^c\,\,\,\,,
%\ee
%where we have used the fact that $\delta_{ab}n^a\dd n^b=\delta_{ab}n^a\bar \dd n^b=0$ and
%$\delta_{ab}n^a\dd\bar \dd n^b=-\delta_{ab}\dd n^a\bar \dd n^b$. Replacing $\chi$ back in the second
%equation we get
%\bea
%{\delta {\cal S}\over \delta n^b} =0=
%\delta_{ba}\left[
%-2B \dd\bar\dd n^a -B^\prime \left(\dd r\bar\dd n^a + \bar \dd r \dd n^a\right)
%- 2 B n^a \delta_{cd}\dd n^c \bar\dd n^d \right] 
%\nonumber \\
%+ \epsilon_{bac}\left[3 \dd n^a \bar \dd n^c + C^\prime n^a\left(\dd r\bar\dd n^c - \bar \dd r \dd n^c\right)
%\right] -3 C \delta_{ba} \epsilon_{cde} n^a n^ c \dd n^d \bar \dd n^e \,\,\,\,.
%\eea
\par
The ansatz for the $SU(2)$ components of the two currents $J=J^aT_a$ and $\bar J=\bar J^aT_a$, appearing in the 
linear system (\ref{J-J}),  is taken 
to have the form \cite{balog1} 
\bea
J^a &=& n^a \dd r + \beta(r)\dd n^a + \gamma(r)\epsilon^{abc}n^b\dd n^c
\nonumber \\
\bar J^a &=& - n^a \bar\dd r - \beta(r)\bar \dd n^a + \gamma(r)\epsilon^{abc}n^b\bar\dd n^c
\eea
Our task now is the determination of the functions $B(r)$, $C(r)$, $\beta(r)$ and 
$\gamma(r)$ such that 
on shell (namely, when the equations of motion (\ref{eom-su(2)}) are obeyed), these two currents satisfy 
\bea
\dd \bar J^a + \bar\dd J^a &=0&
\nonumber \\
\dd \bar J^a - \bar\dd J^a + \epsilon^{abc} J^b\bar J^c &=0&
\label{components}
\eea
In practice, one extracts $\dd\bar\dd r$ and  $\dd\bar\dd n^b$ from
the equations of motion (\ref{eom-su(2)}) and injects them into the above 
on-shell requirements (\ref{components}). 
\par
We find that (\ref{components}) holds provided that the following differential equations 
are obeyed{\footnote{We have used the fact that
$\epsilon_{abc}\,\epsilon_{cde}=\delta_{ad}\delta_{be}-\delta_{ae}\delta_{bd}$
and other relations coming from the differentiation of the constraint
$\delta_{ab}n^an^b=1$. In particular, the relation
$\epsilon^{bcd}n^an^b\dd n^c\bar\dd n^d=\epsilon^{abc}\dd n^b \bar \dd n^c$.}}
\bea
-1 + \beta^\prime - {\gamma\over B}C^\prime=0
\nonumber \\
\gamma^\prime - {\gamma\over B}B^\prime=0
\nonumber \\
-\left(1-\gamma\right) - \beta^\prime  + {\beta\over B}B^\prime=0
\nonumber \\
\gamma^\prime - \beta -{\beta\over B}C^\prime=0
\nonumber \\
2\gamma - \beta^2 - \gamma^2 - {}C^\prime=0
\nonumber \\
2\beta\left(1 - \gamma\right) - {}B^\prime=0
\eea
The first two equations of this system are solved by
\bea
B &=& b\gamma \nonumber \\
C &=& -b\left(r-\beta\right)+c \,\,\,\,\,,
\eea
where $b$ and $c$ are two arbitrary constants, with $b$ different from zero. 
The solutions to the whole system are then 
divided in three cases:
\newline
Case a)
\bea
\gamma=1\,\,\,\,\,\,,\,\,\,\,\,\,\,
\beta=\pm {\sqrt{(1+b)}}
\eea 
with $b$ arbitrary.  
\newline                                                              
Case b)
\bea
b=-{1}\,\,\,\,\,,\,\,\,
\gamma= {\left(r- d\right)^2\over[\left(r-d\right)^2-1]}
\,\,\,\,\,,\,\,\,\,\,
\beta= -{\left(r- d\right)\over[\left(r-d\right)^2-1]}\,\,,
\eea
where $d$ is a constant of integration.
\newline
Case c)
\bea
%&\,& a\equiv {2\sqrt{\left(1+b\right)}\over b}
%\,\,\,\,\,\,\,,\,\,\,\,\,\,
&\,& b\ne - 1\,\,\,\,\,\,\,,\,\,\,\,\,\,
%\eta^2=1\,\,\,\,\,\,\,,\,\,\,\,\,\, 
\nonumber \\
&\,& \gamma={\left[1 + b e^{\eta a(r-d)}\right]^2\over 
\left[1 - be^{\eta a(r-d)}\right]^2-4 e^{\eta a(r-d)}}
\,\,\,\,\,\,\,,\,\,\,\,\,\,
\beta=-{\eta\sqrt{1+b}\left[1-b^2 e^{2\eta a(r-d)}\right]\over 
\left[1 - be^{\eta a(r-d)}\right]^2-4 e^{\eta a(r-d)}} \,\,\,\,\,,
\eea
where $d$ is a constant of integration, $a= {2\over b}{\sqrt{\left(1+b\right)}}$
and $\eta^2=1$.
\par
In summary, the $SU(2)$ integrable non-linear sigma model is given by
\be
S =
\int{\rm d}z{\rm d}\bar z \,
\left\{\dd r\bar \dd r + b\gamma(r)\delta_{ab}\dd n^a\bar\dd n^b
+ \left[b\left(\beta(r)-(r-d)\right) +(c-bd)\right]\epsilon_{abc} n^a\dd n^b\bar \dd n^c \right\}
%+ 
%\chi\left(\delta_{ab}n^an^b -1\right)\right\} 
\,\,\,\,.
\label{su(2)-action}
\ee
It is clear from the expressions of the two functions $\beta(r)$ and $\gamma(r)$ that
the constant $d$ can be absorbed by the change of variable 
$r\longrightarrow r+d$. Furthermore, the term involving the 
constant $(c-bd)$ in the action (\ref{su(2)-action}) is a total derivative
and does not contribute to the equations
of motion.
Therefore, the integrable non-linear sigma model (\ref{su(2)-action}), up to some rescalings of the fields, 
is precisely the model of \cite{balog1}.

%
%
%
%
%
%
%
%
%
%
%\bea
%\gamma={\left(e^{ad} + b e^{ar}\right)^2\over 
%\left(e^{ad} - be^{ar}\right)^2-4 e^{a(r+d)}}
%\,\,\,\,\,\,\,,\,\,\,\,\,\,
%\nonumber \\
%\beta={\sqrt{1+b}\left(b^2 e^{2ar}- e^{2ad}\right)\over 
%\left(e^{ad} - be^{ar}\right)^2-4 e^{a(r+d)}}
%\eea
%Case d)
%\bea
%a\equiv {2\sqrt{\left(1+b\right)}\over b}
%\,\,\,\,\,\,\,,\,\,\,\,\,\,
%\gamma={\left(e^{ar} + be^{ad}\right)^2\over 
%\left(e^{ar} - be^{ad}\right)^2-4 e^{a(r+d)}}
%\,\,\,\,\,\,\,,\,\,\,\,\,\,
%\nonumber \\
%\beta={\sqrt{1+b}\left(b^2 e^{2ad}- e^{2ar}\right)\over 
%\left(e^{ar} - b e^{ad}\right)^2-4 e^{a(r+d)}}
%\eea

\section{Generalisation of the principal chiral model}

We have so far studied only the case when the spectral parameter
enters the Lax pair in a multiplicative manner.
What can one say now about the situations when this is not the case?
Unfortunately, there is not much that one can say about 
a general non-linear sigma model. The integrability is described 
by the Lax pair (\ref{AA}) and the master equation (\ref{F=Dmu})
where one might find a spectral parameter hidden in the matrices
$\alpha_i$ and $\beta_i$.
However, a great deal can be learnt by exploring particular
non-linear sigma models. 
\par
We start by studying a theory which generalises the 
principal chiral non-linear sigma model. The
classical integrability of this model has already
been investigated in the literature \cite{sochen,hlavaty}.
Its study here is for two purposes: Firstly, we would like to
highlight the group theory structure behind the integrability requirement. 
Secondly, the results of this example will be of use in the rest of the 
paper.  
\par
We consider the generalisation of the principal chiral non-linear sigma model
as given by the action
\be
S\left(g\right)=\int{\rm d}z{\rm d}\bar z \,
\Omega_{ab}\left(g^{-1}\dd g\right)^a
\left(g^{-1}\bar\dd g\right)^b
\,\,\,\,,
\label{gen-pcm}
\ee
where $\Omega_{ab}$ is a constant matrix
and $g^{-1}\dd g=\left(g^{-1}\dd g\right)^aT_a=A^aT_a$, 
$g^{-1}\bar\dd g=\left(g^{-1}\bar\dd g\right)^aT_a=\bar A^aT_a$
and $T_a$ are some Lie algebra generators, 
$\left[T_a\,,\,T_b\right]=f^c_{ab}T_c$,
and $g$ an element
in the corresponding Lie group. 
The equations 
of motion are conveniently expressed as 
\bea
{\cal{E}}_c\equiv -{1\over 2}\left(\Omega_{cd}+\Omega_{dc}\right)
\left(\dd \bar A^d + \bar \dd A^d\right)
+\left[{1\over 2}\left(\Omega_{cd}-\Omega_{dc}\right)f^d_{ab}
+\left(\Omega_{ad}f^d_{bc} + \Omega_{db}f^d_{ac}\right)\right]A^a\bar A^b =0\,,
\label{eom-original}
\eea
where use of the Bianchi identities
\bea
{\cal{B}}^a\equiv \dd \bar A^a - \bar \dd A^a
+ f^a_{bc} A^b\bar A^c =0\,\,\,\,,
\label{bianchi-original}
\eea
has been made.
\par
The linear system, whose consistency condition is equivalent to the above
equations of motion, could only be of the form
\bea
&&\left(\dd + A^a P_a\right)\Psi=0
\nonumber\\
&&\left(\bar\dd + \bar A^b Q_b\right)\Psi=0
\,\,\,\,\,\,,
\label{gen-PCS-lax}
\eea
where $P_a$ and $Q_a$ are constant matrices. The spectral parameter, if it exists,
is hidden in these matrices. The compatibility condition of this linear system is
\bea
{\cal F}\equiv 
{1\over 2}\left(Q_a-P_a\right)\left(\dd\bar A^a+\bar\dd A^a\right)
+\left(-{1\over 2}f^d_{ab}\left(Q_d+P_d\right)+\left[P_a\,,\,Q_b\right]\right)
A^a\bar A^b =0\,\,\,\,.
\eea
We then demand that 
\be
{\cal F}=R^a{\cal E}_a\,\,\,\,
\ee
for some linearly independent matrices $R^a$. This leads  
firstly  to the determination 
of the matrices $P_a$ in terms of $Q_a$ and $R^a$
through 
\bea
P_a=Q_a+(\Omega_{ac}+\Omega_{ca})R^c\,\,\,\,. 
\eea
Secondly, it yields
the condition
\bea
%P_a &=&Q_a+(\Omega_{ac}+\Omega_{ca})R^c
%\nonumber \\
\left[Q_a\,,\,Q_b\right] =f^d_{ab}Q_d +\Omega_{cd}f^d_{ab}R^c+
\left(\Omega_{ad}f^d_{bc}+\Omega_{db}f^d_{ac}\right)R^c
-\left(\Omega_{ac}+\Omega_{ca}\right)\left[R^c\,,\,Q_b\right]
\label{genPCS}
\eea
This last equation determines the non-linear sigma model (that is $\Omega_{ab}$)
and its Lax pair (the matrices $Q_a$ and $R^a$). 
It is clear that if a solution exists, then it must have a 
Lie algebra interpretation. We will show below that equation (\ref{genPCS})
does have solutions for particular examples.
%\par
%It is also instructive to examine the case for which
%$\Omega_{ab}$ is symmetric. Let us introduce the matrices $X_a$ through 
%the definition $Q_a=X_a-R_a$, where $R_a=\Omega_{ab}R^c$ (in this case 
%$P_a=X_a+R_a$). The anti-symmetric and symmetric parts of equation (\ref{genPCS})
%yield then
%\bea
%\left[X_a\,,\,X_b\right] - \left[R_a\,,\,R_b\right] &=& f^c_{ab}X_c
%\nonumber \\
%\left[X_a\,,\,R_b\right]+\left[X_b\,,\,R_a\right] &=& 
%-\left(\Omega_{ac}f^c_{bd}+\Omega_{bc}f^c_{ad}\right)\Omega^{de}R_e\,\,\,\,\,,
%\eea
%where $\Omega^{ab}$ is the inverse of $\Omega_{ab}$.

\noindent
\vskip1.0cm
\noindent
{\it{\Large{A three-dimensional example}}}
\vskip1.0cm

This particular solution to (\ref{genPCS}) is based on 
the $SU(2)$ Lie algebra. We take the matrix $\Omega_{ab}$
to be diagonal
\be
\Omega_{ab}={\rm diag}\left(L_1\,,\,L_2\,,\,L_3\right)\,\,\,\,.
\label{su2-omega}
\ee
The linear system is taken to have the form
\bea
&&\left(\dd + X^a_b A^b T_a\right)\Psi=0
\nonumber\\
&&\left(\bar\dd + Y^c_d \bar A^d T_c\right)\Psi=0
\,\,\,\,\,\,,
\label{su2-lax}
\eea
where the $SU(2)$ Lie algebra is $\left[T_a\,,\,T_b\right]=\epsilon_{abc}T_c$
with $\epsilon_{123}=1$ and  $X^a_b$ and $Y^a_b$ are constant 
$3\times 3$ matrices.
In the notation of (\ref{gen-PCS-lax}), we have 
$P_a=X_a^bT_b$, $Q_a=Y_a^bT_b$ and $R^d={1\over 2}(\Omega^{-1})^{da}(X^b_a-Y^b_a)T_b$.
\par
It is then found that equation (\ref{genPCS}) is satisfied for 
the following non-vanishing elements $X^a_b$ and $Y^a_b$:
\bea
X^1_1 &=& {1\over \sqrt{L_2}\sqrt{L_3}}
\left[\kappa_1\sqrt{x+L_{2}}\sqrt{x+L_{3}}
+\omega_1\sqrt{x}\sqrt{x+L_{1}}\right]
\nonumber\\
X^2_2 &=& {1\over \sqrt{L_1}\sqrt{L_3}}
\left[\kappa_2\sqrt{x+L_{1}}\sqrt{x+L_{3}}
+\omega_2\sqrt{x}\sqrt{x+L_{2}}\right]
\nonumber\\
X^3_3 &=& {1\over \sqrt{L_1}\sqrt{L_2}}
\left[\kappa_3\sqrt{x+L_{1}}\sqrt{x+L_{3}}
+\omega_3\sqrt{x}\sqrt{x+L_{3}}\right]
\nonumber\\
Y^1_1 &=& {1\over \sqrt{L_2}\sqrt{L_3}}
\left[\kappa_1\sqrt{x+L_{2}}\sqrt{x+L_{3}}
-\omega_1\sqrt{x}\sqrt{x+L_{1}}\right]
\nonumber\\
Y^2_2 &=& {1\over \sqrt{L_1}\sqrt{L_3}}
\left[\kappa_2\sqrt{x+L_{1}}\sqrt{x+L_{3}}
-\omega_2\sqrt{x}\sqrt{x+L_{2}}\right]
\nonumber\\
Y^3_3 &=& {1\over \sqrt{L_1}\sqrt{L_2}}
\left[\kappa_3\sqrt{x+L_{1}}\sqrt{x+L_{3}}
-\omega_3\sqrt{x}\sqrt{x+L_{3}}\right]\,\,\,\,,
\label{hlavaty}
\eea
where $\kappa_1^2=\kappa_2^2=1$, $\kappa_3=\kappa_1\kappa_2$
and $\omega_1=\kappa_2\omega_3$, $\omega_2=\kappa_1\omega_3$,
$\omega_3^2=1$. The spectral parameter is $x$.
This solution was first found in \cite{cherednik} and later given in its 
present form in \cite{hlavaty}. 
\par
This solution might as well be extended to the case of the $SO(6)$ Lie algebra which is 
described  by the commutations relations 
\bea
\left[J_a\,,\,J_b\right]=\epsilon_{abc}J_c\,\,\,\,,\,\,\,\,
\left[J_a\,,\,M_b\right]=\epsilon_{abc}M_c\,\,\,\,,\,\,\,\,
\left[M_a\,,\,M_b\right]=\epsilon_{abc}J_c\,\,\,\,.\,\,\,\,
\eea
However, this Lie algebra can be cast in the form
\bea
\left[J^{\pm}_a\,,\,J^{\pm}_b\right]=\epsilon_{abc}J^{\pm}_c\,\,\,\,\,,\,\,\,\,\,\left[J^{+}_a\,,\,J^{-}_b\right]=0
\,\,\,\,\,\,\,\,{\rm{with}}\,\,\,\,\,\,\,\,
 J_a^{\pm}={1\over 2}\left(J_a\pm M_a\right)\,\,\,\,.
\eea
We have therefore two copies of the $SU(2)$ Lie algebras and we conclude that 
the non-linear sigma model (\ref{gen-pcm}) defined by  
\be
\Omega_{ab}={\rm diag}\left(L^+_1\,,\,L^+_2\,,\,L^+_3\,,\,L_1^-\,,\,L_2^-\,,\,L^-_3\right)\,\,\,\,
\ee
is integrable. The corresponding linear system is of the form (\ref{su2-lax})
where now the indices $a\,,\,b\,\dots$ run from $1$ to $6$.
The matrix elements $\left(X^1_1\,, \,X^2_2\,,\,X^3_3\right)$ 
and $\left(X^4_4\,, \,X^5_5\,,\,X^6_6\right)$ 
are obtained from (\ref{hlavaty}) by replacing 
$(L_1\,,\,L_2\,,\,L_3)$ by $(L_1^+\,,\,L_2^+\,,\,L_3^+)$ and $(L_1^-\,,\,L_2^-\,,\,L_3^-)$,
respectively. The determination of $Y^a_b$ is similar.
%\bea
%&&\left[\dd + \left(X^a_{+b} J^+_a + X^a_{-b}J^-_a\right)\left(A^b_++A^b_-\right) \right]\Psi=0
%\nonumber\\
%&&\left[\bar\dd + \left(Y^a_{+b} J^+_a + Y^a_{-b}J^-_a\right)\left(\bar A^b_++\bar A^b_-\right) \right]\Psi=0
%\,\,\,\,\,\,,
%\label{so6-lax}
%\eea
%where $A=g^{-1}\bar\dd g=A^a_+J^+_a+A^a_-J^-_a$ and 
%$\bar A=g^{-1}\bar\dd g=\bar A^a_+J^+_a+\bar A^a_-J^-_a$. The matrices 
%$X^a_{+b}$ and $Y^c_{+d}$ are obtained from the expressions of $X^a_{b}$ and $Y^c_{d}$
%in (\ref{hlavaty}) upon replacing, respectively, $(L_1\,,\,L_2\,,\,L_3)$ by $(L_1^+\,,\,L_2^+\,,\,L_3^+)$,
%while $X^a_{-b}$ and $Y^c_{-d}$ are obtained by 
%substituting $(L_1\,,\,L_2\,,\,L_3)$ with $(L_1^-\,,\,L_2^-\,,\,L_3^-)$.

\noindent
\vskip1.0cm
\noindent
{\it{\Large{A four-dimensional example}}}
\vskip1.0cm

So far, all the integrable non-linear sigma models listed in this paper are known.
We will now explicitly construct a new integrable model. This is based on 
a four-dimensional non-semi-simple Lie algebra whose commutation relations are
\bea
\left[J\,,\,P_i\right]=\epsilon_{ij}P_j\,\,\,\,,\,\,\,\,
\left[P_i\,,\,P_j\right]=\epsilon_{ij}T,\,\,\,,\,\,\,\,
\left[T\,,\,J\right]=\left[T\,,\,P_i\right]=0,\,\,\,
\label{NW-algebra}
\eea
with $\epsilon_{12}=1$. We denote the generators of this Lie algebra
as $T_a=\left\{P_1\,,\,P_2\,,\,J\,,\,T\right\}$ such that the
Lie algebra is $\left[T_a\,,\,T_b\right]=f^c_{ab}\,T_c$. Although this algebra
is non-semi-simple, it has nevertheless a non-degenerate bilinear form
$\eta_{ab}$ obeying $\eta_{ab}f^b_{cd}+\eta_{cb}f^b_{ad}=0$. It 
is given by \cite{witten-nappi}
\bea
\eta_{ab}=\left(
\begin{array}{cccc}
1 & 0& 0& 0 \\
0 &1 & 0& 0 \\
0 &0 & b& 1 \\
0 &0 & 1& 0
\end{array} \right)\,\,\,\,\,,
\label{NW-form}
\eea
where $b$ is an arbitrary constant. 
\par
Using this invertible bilinear form
one can, for instance,  construct a principal chiral non-linear sigma model 
\be
S\left(g\right)=\int{\rm d}z{\rm d}\bar z \,
\eta_{ab}\left(g^{-1}\dd g\right)^a
\left(g^{-1}\bar\dd g\right)^b
\,\,\,\,,
\label{PC-nappi-witten}
\ee
where as usual $\left(g^{-1}\dd g\right)= \left(g^{-1}\dd g\right)^aT_a=A^a T_a$ and 
$\left(g^{-1}\bar\dd g\right)=\left(g^{-1}\bar\dd g\right)^aT_a=\bar A^aT_a$. This is an integrable
non-linear sigma model with the Lax pair
\bea
\left[\dd +{1\over 1+x}\left(g^{-1}\dd g\right)\right]\Psi &=&0
\nonumber\\
\left[\bar\dd +{1\over 1-x}\left(g^{-1}\bar\dd g\right)\right]\Psi &=&0\,\,\,\,,
\eea
where $x$ is the spectral parameter.
\par
Our aim is to generalise the above principal chiral non-linear sigma model.
We consider the action
\be
S\left(g\right)=\int{\rm d}z{\rm d}\bar z \,
\Omega_{ab}\left(g^{-1}\dd g\right)^a
\left(g^{-1}\bar\dd g\right)^b
\,\,\,\,,
\ee
where $\Omega_{ab}$ is symmetric and of the form
\bea
\Omega_{ab}=\left(
\begin{array}{cccc}
L_{1} & 0& 0& 0 \\
0 &L_{2} & 0& 0 \\
0 &0 & b& L_{3} \\
0 &0 & L_{3}& 0
\end{array} \right)\,\,\,\,\,.
\label{4-d}
\eea
This non-linear sigma model is integrable if a solution to 
(\ref{genPCS}) can be found. {}For this purpose, we seek a linear 
system of the form
\bea
&&\left(\dd + X^a_b A^b T_a\right)\Psi=0
\nonumber\\
&&\left(\bar\dd + Y^c_d \bar A^d T_c\right)\Psi=0
\,\,\,\,\,\,,
\label{witten-nappi-lax}
\eea
where we assume that the constant matrices $X^a_b$ and  $Y^c_d$ are diagonal
(that is, eight unknowns).
It turns out that $(\ref{genPCS})$ yields six independent equations (for eight unknowns).
These are 
\bea
-2L_3Y^2_2X_1^1+\left(L_1-L_2\right)\left(X^4_4-Y^4_4\right)+L_3\left(X^4_4+Y^4_4\right)&=& 0
\nonumber \\
+2L_3Y^1_1X_2^2+\left(L_1-L_2\right)\left(X^4_4-Y^4_4\right)-L_3\left(X^4_4+Y^4_4\right)&=& 0
\nonumber \\
+2L_2Y^3_3X_1^1+\left(L_3-L_1\right)\left(X^2_2-Y^2_2\right)-L_2\left(X^2_2+Y^2_2\right)&=& 0
\nonumber \\
+2L_2Y^1_1X_3^3+\left(L_1-L_3\right)\left(X^2_2-Y^2_2\right)-L_2\left(X^2_2+Y^2_2\right)&=& 0
\nonumber \\
-2L_1Y^3_3X_2^2+\left(L_2-L_3\right)\left(X^1_1-Y^1_1\right)+L_1\left(X^1_1+Y^1_1\right)&=& 0
\nonumber \\
-2L_1Y^2_2X_3^3+\left(L_3-L_2\right)\left(X^1_1-Y^1_1\right)+L_1\left(X^1_1+Y^1_1\right)&=& 0 \,\,\,\,.
\label{witten-nappi-set}
\eea
The explicit solution to this set is too untidy to write down. We choose therefore to give it in an iterative 
manner. 
\par
We take the two elements $X^1_1$ and $Y^1_1$ as our free parameters leaving us with
six equations and six unknowns. One of these parameters
reflects simply the fact that our Lie algebra is invariant under the rescaling
$P_1\rightarrow \lambda P_1$, $P_2\rightarrow \lambda P_2$, $J\rightarrow J$, $T\rightarrow \lambda^2 T$.
Therefore, only one parameter is left
and in order to simplify slightly the equations we take
$X^1_1=1+x$ and $Y^1_1=1-x$. The parameter $x$ will play the role of the spectral parameter.
The last two equations of (\ref{witten-nappi-set}) yield then
\bea
X^3_3  &=&{1\over Y^2_2}\left[1-{x\left(L_2-L_3\right)\over L_1}\right]
\nonumber \\
Y^3_3  &=&{1\over X^2_2 }\left[1+{x\left(L_2-L_3\right)\over L_1}\right] \,\,\,\,
\eea
while the first and the second equations give
\bea
X^4_4 &=&  
{\left(L_1-L_2-L_3\right)\left(1-x\right)\over 2\left(L_1-L_2\right)} X^2_2 +
        {\left(L_1 -L_2+L_3\right)\left(1+x\right)\over 2\left(L_1-L_2\right)} Y^2_2 
\nonumber \\	
Y^4_4 &=&  
{\left(L_1-L_2+L_3\right)\left(1-x\right)\over 2\left(L_1-L_2\right)} X^2_2 +
        {\left(L_1 -L_2-L_3\right)\left(1+x\right)\over 2\left(L_1-L_2\right)}Y^2_2 \,\,\,\,.	
\eea
Replacing these back into (\ref{witten-nappi-set}) allows the determination of $Y^2_2$ 
\bea
Y^2_2={\left(L_3-L_1 -L_2\right)\over \left(L_3 -L_1+L_2\right)}X^2_2
+{2\left(1+x\right)L_2 \left(L_1+x\left(L_2-L_3\right)\right)\over L_1\left(L_3 -L_1+L_2\right)}
{1\over X^2_2}\,\,\,\,.
\eea
Finally, $X^2_2$ is obtained by solving the equation
\bea
&L_1^2\left(L_1-L_3\right)\left(X^2_2\right)^4
-L_1\left[\left(L_3-L_1-L_2\right)\left(\left(L_3-L_2\right)x^2
-L_1\right)
+4L_2\left(L_1-L_3\right)x
\right]\left(X^2_2\right)^2 &
\nonumber \\
&+L_2\left(1+x\right)^2\left(L_1+x\left(L_2-L_3\right)\right)^2=0 &
\eea
This is a quadratic equation in $z=(X^2_2)^2$ whose solution is
\bea
\left(X^2_2\right)^2 &=&{1\over 2 L_1^2\left(L_1-L_3\right)}
\left\{
L_1\left[\left(L_3-L_1-L_2\right)\left(\left(L_3-L_2\right)x^2
-L_1\right)
+4L_2\left(L_1-L_3\right)x
\right] \right.
\nonumber \\
&\pm& 
\left\{\left[L_1\left[\left(L_3-L_1-L_2\right)\left(\left(L_3-L_2\right)x^2
-L_1\right)
+4L_2\left(L_1-L_3\right)x
\right]\right]^2 \right.
\nonumber \\
&-& \left.
4L_1^2L_2\left(L_1-L_3\right)\left(1+x\right)^2\left(L_1+x\left(L_2-L_3\right)\right)^2\right\}^{1/2}
\left.\right\}\,\,\,\,\,.
\eea
Once $X^2_2$ has been determined all the other elements $X^a_a$ and $Y^a_a$ are found. 
\par
We should mention that this solution is not valid for  
the two values of the spectral parameter $x=-1$ and $x=L_1/(L_3-L_2)$ as 
they lead to $X^2_2=0$. Furthermore, there are special cases corresponding to 
particular values of $L_1$, $L_2$ and $L_3$ which are not discussed here.
These particular cases are such that $L_1-L_2=0$, $L_1-L_3=0$, $L_3 -L_1+L_2=0$ or $L_3 -L_1-L_2=0$. 

\noindent
\vskip1.0cm
\noindent
{\it{\Large{A five-dimensional example}}}
\vskip1.0cm

We consider a five-dimensional non-semi-simple Lie algebra whose generators
$T_a=\left\{P_1\,,\,P_2\,,\,P_3\,,\,J\,,\,T\right\}$ 
satisfy $\left[T_a\,,\,T_b\right]=f^c_{ab}\,T_c$ with the non-vanishing 
commutators being
\bea
\left[P_i\,,\,P_j\right]=\epsilon_{ijk}\,P_k +v^k\,\epsilon_{kij}\,J\,\,\,\,\,\,,\,\,\,\,\,\,\,
\left[T\,,\,P_i\right]=v^k\,\epsilon_{kij}\,P_j\,\,\,\,\,\,\,,
\eea
where $\epsilon_{123}=1$ an $v^k$ is an arbitrary constant vector.
\par
This Lie algebra possesses an invertible bilinear form given by \cite{jose-sonia}
\bea
\eta_{ab}=\left(
\begin{array}{ccccc}
1 & 0& 0& 0 &0 \\
0 &1 & 0& 0 &0 \\
0 &0 & 1& 0  & 0\\
0 &0 & 0 & 0 &1 \\
0 &0 & 0 & 1 &0
\end{array} \right)\,\,\,\,\,.
\label{jose's-eta}
\eea
Of course, as in the above four-dimensional example, the principal chiral model constructed with
this bilinear form is integrable. 
\par
The integrable non-linear sigma model we present here is described by the
action
\be
S\left(g\right)=\int{\rm d}z{\rm d}\bar z \,
\Omega_{ab}\left(g^{-1}\dd g\right)^a
\left(g^{-1}\bar\dd g\right)^b
\,\,\,\,,
\ee
where $\Omega_{ab}$ is a slight generalisation of the bilinear form (\ref{jose's-eta})
and is of the form
\bea
\Omega_{ab}=\left(
\begin{array}{ccccc}
1 & 0& 0& 0 &0\\
0 &1 & 0& 0 &0\\
0 &0 & 1& 0 &0 \\
0 &0 & 0& 0 & L \\
0 &0 & 0& L & 0
\end{array} \right)\,\,\,\,\,.
\label{5-d}
\eea
This non-linear sigma model is integrable and its corresponding linear 
system is given by{\footnote{The condition of integrability (\ref{genPCS})
leads to a long list of equations which we will not write down. Here we will only give 
the result of a computer based investigation.}}
\bea
&&\left(\dd + X^a_b A^b T_a\right)\Psi=0
\nonumber\\
&&\left(\bar\dd + Y^c_d \bar A^d T_c\right)\Psi=0
\,\,\,\,\,\,,
\label{jose's-lax}
\eea
where the constant matrices $X^a_b$ and  $Y^c_d$ are 
\bea
X^a_b &=& \left(
\begin{array}{ccccc}
x & 0& 0& 0 &0 \\
0 &x & 0& 0 &0 \\
0 &0 & x& 0  & 0\\
0 &0 & 0 & y &0 \\
0 &0 & 0 & 0 &1+L\left(x-1\right)
\end{array} \right)\,\,\,\,\,
\nonumber \\
Y^a_b &=& {1\over 2x-1}\left(
\begin{array}{ccccc}
{x} & 0& 0& 0 &0 \\
0 & {x} & 0& 0 &0 \\
0 &0 & {x} & 0  & 0\\
0 &0 & 0 & 2x^2-y\left(2x-1\right) &0 \\
0 &0 & 0 & 0 &\left(2x-1\right)-L\left(x-1\right)
\end{array} \right)\,\,\,\,\,.
\eea
The parameters $x$ and $y$ are arbitrary.

\noindent
\vskip1.0cm
\noindent
{\it{\Large{Klim\v{c}\'{\i}k's solution}}}{\footnote{This solution was published during 
the reviewing process of the present paper.}}
\vskip1.0cm

%In order to introduce this solution, let us first start by establishing some 
%useful definitions. A Lie algebra ${\cal G}$ is generated by 
%$(H^\mu\,,\,B^{\alpha}\,,\, C^{\alpha})$ where $\alpha$ labels the positive roots
%and $H^\mu$ are the generators of the Cartan subalgebra. 
%The step generators $E^{\alpha}$ and $E^{-\alpha}$ are defined as
%\be
%E^{\alpha}=B^{\alpha}+iC^{\alpha} \,\,\,,\,\,\,
%E^{-\alpha}=B^{\alpha}-iC^{\alpha}\,\,\,\,.
%\ee
%The most important ingredient in the construction of [] is the linear
%operator $R$ which acts on the generators of the Lie algebra ${\cal G}$ as follows:
%\be
%RH^\mu=0\,\,\,\,,\,\,\,\,RB^\alpha=C^\alpha\,\,\,\,,\,\,\,\, 
%RC^\alpha=-B^\alpha\,\,\,\,.
%\ee
%{}For instance, in the case of the Lie algebra $SU(2)$ we have
%\bea
%R\vec{\sigma}=\left(
%\begin{array}{ccc}
%0 & 1 & 0\\
%-1 & 0 & 0 \\
%0& 0& 0
%\end{array} \right) 
%\left(\begin{array}{c}
%\sigma_1\\
%\sigma_2\\
%\sigma_3
%\end{array}\right)= \left(\begin{array}{c}
%\sigma_2\\
%-\sigma_1\\
%0
%\end{array}\right)\,\,\,\,,
%\eea
%where $\sigma_i$ are the usual Pauli matrices. Similarly, for ${\cal G}=SU(3)$ the operator $R$ acts as
%\bea
%&& R\lambda_1=\lambda_2\,\,,\,\,R\lambda_2=-\lambda_1\,\,,\,\,R\lambda_3=0\,\,,\,\,
%R\lambda_4=\lambda_5\,\,,\,\,
%\nonumber \\
%&&R\lambda_5=-\lambda_4\,\,,\,\, R\lambda_6=\lambda_7\,\,,\,\,R\lambda_7=-\lambda_6\,\,,\,\,
%R\lambda_8=0\,\,\,,
%\eea
%where $\lambda_i$ are the Gell-Mann matrices.
%\par
The most important ingredient in the construction of \cite{klimcik3} is a linear
operator $R$ acting on the generators of a simple Lie algebra ${\cal G}$.
This operator is required to satisfy two relations. Firstly 
\bea
\left[RX\,,\,RY\right] = R\left(\left[X\,,\,Y\right]_R\right)+\left[X\,,\,Y\right]\,\,\,\,\,\,\,{\rm {for}}\,\,\,\,\,\,\,
X\,,\,Y \,\,\in {\cal G}\,\,\,\,,
\label{R-1}
\eea
where
\bea
\left[X\,,\,Y\right]_R &\equiv & \left[RX\,,\,Y\right]+\left[X\,,\,RY\right]\,\,\,\,.
\eea
Secondly, $R$ verifies the
skew-symmetry condition
\be
<RX\,,\,Y>_{{\cal G}}+ <X\,,\,RY>_{{\cal G}}=0\,\,\,\,,
\label{R-2}
\ee
where $<\,,\,>_{{\cal G}}$ is the Killing-Cartan form on the Lie algebra ${\cal G}$. 
\par
An $R$ operator satisfying (\ref{R-1}) and (\ref{R-2}) was given in \cite{klimcik3}. This is 
constructed as follows: 
Let the Lie algebra ${\cal G}$ be generated by 
$(H^\mu\,,\,B^{\alpha}\,,\, C^{\alpha})$ where $\alpha$ labels the positive roots
and $H^\mu$ are the generators of the Cartan subalgebra.  
The step generators $E^{\alpha}$ and $E^{-\alpha}$ are defined, up to a normalisation factor, as
\be
E^{\alpha}=B^{\alpha}+iC^{\alpha} \,\,\,,\,\,\,
E^{-\alpha}=B^{\alpha}-iC^{\alpha}\,\,\,\,.
\ee
The linear
operator $R$ acts on the generators of the Lie algebra ${\cal G}$ as follows:
\be
RH^\mu=0\,\,\,\,,\,\,\,\,RB^\alpha=C^\alpha\,\,\,\,,\,\,\,\, 
RC^\alpha=-B^\alpha\,\,\,\,.
\label{R-3}
\ee
{}For instance, in the case of the Lie algebra $SU(2)$, with generators $\vec T=\{T_1\,,\,T_2\,,\,T_3\}$, we have
\bea
R\vec{T}=\left(
\begin{array}{ccc}
0 & 1 & 0\\
-1 & 0 & 0 \\
0& 0& 0
\end{array} \right) 
\left(\begin{array}{c}
T_1\\
T_2\\
T_3
\end{array}\right)= \left(\begin{array}{c}
T_2\\
-T_1\\
0
\end{array}\right)\,\,\,\,,
\eea
Similarly, for ${\cal G}=SU(3)$ the operator $R$ acts as
\bea
&& RT_1=T_2\,\,,\,\,RT_2=-T_1\,\,,\,\,RT_3=0\,\,,\,\,
RT_4=T_5\,\,,\,\,
\nonumber \\
&&RT_5=-T_4\,\,,\,\, RT_6=T_7\,\,,\,\,RT_7=-T_6\,\,,\,\,
RT_8=0\,\,\,.
\eea
We will show below that (at least for the $SU(2)$ case) the above action of 
the operator $R$ is not the most general one.
\par
The integrable non-linear sigma model constructed in \cite{klimcik3} is described by the action
\be
S\left(g\right)=\int{\rm d}z{\rm d}\bar z \,
<g^{-1}\dd g\,\,,\,\,
\left(I-\varepsilon R\right)^{-1}g^{-1}\bar\dd g>_{{\cal G}}
\,\,\,\,,
\label{klimcik}
\ee
where $g$ is the group element corresponding to the Lie algebra ${\cal G}$, $I$ is the identity
operation on the generators of ${\cal G}$ and $\varepsilon$ is a free parameter. The equations
of motion of this theory can be cast into the form
\be
{\cal E} \equiv \dd \bar J -\bar \dd J + \varepsilon \left[J\,,\,\bar J\right]_R=0\,\,\,\,,
\ee
where the two currents $J$ and $\bar J$ are defined as
\be
J = \left(I+\varepsilon R\right)^{-1}g^{-1} \dd g\,\,\,\,,\,\,\,\,
\bar J = -\left(I-\varepsilon R\right)^{-1}g^{-1}\bar\dd g \,\,\,\,.
\ee
These two currents satisfy the Bianchi identity
\bea
{\cal B} &\equiv&  -\left(\dd \bar J +\bar \dd J\right) 
+\varepsilon R\left(\dd \bar J -\bar \dd J + \varepsilon \left[J\,,\,\bar J\right]_R\right)
+\left(\varepsilon^2-1\right)\left[J\,,\,\bar J\right]
\nonumber \\
 &-&\varepsilon\left[RJ\,,\,\bar J\right]
+\varepsilon\left[J\,,\,\bar RJ\right] \,\,\,\,.
\eea
These are found by writing $g^{-1} \dd g=\left(I+\varepsilon R\right)J$
and $g^{-1}\bar\dd g= -\left(I-\varepsilon R\right)\bar J$ and demanding that
the identity $\dd\left(g^{-1} \bar\dd g\right) -\bar\dd\left(g^{-1}\dd g\right)
+\left[g^{-1} \dd g \,,\,g^{-1}\bar\dd g\right]=0$ holds.
\par
Finally, the Lax pair corresponding to this non-linear sigma is given by
\bea
\left[\dd -\left(\varepsilon^2 -\varepsilon R -{1+\varepsilon^2\over 1+x}\right)J\right]\Psi &=& 0 
\nonumber \\
\left[\bar \dd +\left(\varepsilon^2 +\varepsilon R -{1+\varepsilon^2\over 1-x}\right)\bar J\right]\Psi &=& 0
\,\,\,\,
\label{lax-klimcik}
\eea
with $x$ being the spectral parameter. The zero curvature condition of this linear system
is
\bea
{\cal F} =-{x\left(1+\varepsilon^2\right)\over 1-x^2}\left(1+x\varepsilon R\right){\cal E}
+{1+x^2\varepsilon^2\over 1-x^2}\,{\cal B}\,\,\,\,.
\eea
We see that the second term vanishes identically while the first  is equivalent to the 
equations of motion.
\par
Of course the action (\ref{klimcik}) is a particular case of the generalised 
chiral principal model (\ref{gen-pcm}). {}For example, in the case of the Lie
algebra $SU(2)$, the model (\ref{klimcik}) leads to
\bea
\Omega_{ab}={1\over 1+\varepsilon^2}\left(
\begin{array}{ccc}
1&-\varepsilon& 0\\
\varepsilon& 1 & 0\\
0&0&1+\varepsilon^2
\end{array} \right)\,\,\,\,.
\label{klimcik-su2}
\eea
The anti-symmetric part of $\Omega_{ab}$ yields a total derivative in the action and one is left
with the integrable model of Cherednik \cite{cherednik}.  {}For completeness, we also give 
the matrices $P_a=X^b_aT_b$ and $Q_a=Y^b_aT_b$ appearing in the linear system (\ref{gen-PCS-lax})
by comparison with (\ref{lax-klimcik})
\bea
X^b_a={1\over 1+x}\left(
\begin{array}{ccc}
1&x\varepsilon& 0\\
-x\varepsilon& 1 & 0\\
0&0&1-x\varepsilon^2
\end{array} \right)\,\,\,\,,\,\,\,\,\,
Y^b_a={1\over 1-x}\left(
\begin{array}{ccc}
1&x\varepsilon& 0\\
-x\varepsilon& 1 & 0\\
0&0&1+x\varepsilon^2
\end{array} \right)\,\,\,\,,\,\,\,\,\,
\eea
where the lower index labels the rows.
\par
Similarly, the tensor $\Omega_{ab}$ corresponding to the $SU(3)$ case is given by
\bea
\Omega_{ab}={1\over 1+\varepsilon^2}\left(
\begin{array}{cccccccc}
1          &-\varepsilon& 0 &0 &0 &0 &0 &0 \\
\varepsilon&         1  & 0 &0 &0 &0 &0 &0\\
0          &0           &1+\varepsilon^2 &0 &0 &0 &0 &0\\
0& 0& 0& 1 &-\varepsilon& 0 &0 &0\\
0& 0& 0 &\varepsilon& 1 &0 &0 &0\\
0& 0& 0& 0 & 0& 1 &-\varepsilon& 0 \\
0& 0& 0& 0 & 0& \varepsilon& 1&  0 \\
0& 0& 0& 0& 0 &0 & 0& 1 +\varepsilon^2 
\end{array} \right)\,\,\,\,.
\eea
Here the anti-symmetric part does not result in a total derivative in the action. 
The explicit form of the $SU(3)$ matrices 
$P_a=X^b_aT_b$ and $Q_a=Y^b_aT_b$ of the linear system (\ref{gen-PCS-lax})
are
\bea
X_a^b={1\over 1+x}\left(
\begin{array}{cccccccc}
1          &x\varepsilon& 0 &0 &0 &0 &0 &0 \\
-x\varepsilon&         1  & 0 &0 &0 &0 &0 &0\\
0          &0           &1-x\varepsilon^2 &0 &0 &0 &0 &0\\
0& 0& 0& 1 &x\varepsilon& 0 &0 &0\\
0& 0& 0 &-x\varepsilon& 1 &0 &0 &0\\
0& 0& 0& 0 & 0& 1 &x\varepsilon& 0 \\
0& 0& 0& 0 & 0& -x\varepsilon& 1&  0 \\
0& 0& 0& 0& 0 &0 & 0& 1 -x\varepsilon^2 
\end{array} \right)\,\,\,\,.
\eea
The matrix $Y_a^b$ is obtained from the expression of $X^b_a$ by replacing
$x$ with $-x$ and $\varepsilon$ with $-\varepsilon$.
\par
As mentioned above, it seems that the action of the operator $R$ as
given in (\ref{R-3}) is not the most general one. Indeed, for the 
case of the Lie algebra $SU(2)$ we have found that the most general
linear operator $R$, satisfying the two conditions (\ref{R-1}) and (\ref{R-2}),  is given by
\bea
R\vec{T}=\left(
\begin{array}{ccc}
0 & a & c\\
-a & 0 & b \\
-c & -b& 0
\end{array} \right) 
\left(\begin{array}{c}
T_1\\
T_2\\
T_3
\end{array}\right)
%= \left(\begin{array}{c}
%T_2\\
%-T_1\\
%0
%\end{array}\right)
\,\,\,\,,
\eea
where $a$ and $b$ are two arbitrary parameters and $c=\sqrt{1-a^2-b^2}$. In this case, the action (\ref{klimcik}) yields  
an integrable non-linear sigma model of the form (\ref{gen-pcm}) with 
\bea
\Omega_{ab}={1\over 1+ \varepsilon^2}\left(
\begin{array}{ccc}
\left(1+\varepsilon^2b^2\right) & -\varepsilon\left(a +\varepsilon bc\right) &
\varepsilon\left(\varepsilon a b -c\right) \\
\varepsilon\left(a -\varepsilon bc\right) & 1+\varepsilon^2c^2 &
-\varepsilon\left(b +\varepsilon ac\right) \\
\varepsilon\left(\varepsilon a b +c\right) & \varepsilon\left(b -\varepsilon ac\right) &
\left(1+\varepsilon^2a^2\right)
\end{array}
\right)\,\,\,\,.
\label{my-su2}
\eea
The corresponding Lax pair  as read from (\ref{lax-klimcik}) results in a linear system of 
the form (\ref{gen-PCS-lax}) with  $P_a=X^b_aT_b$ and $Q_a=Y^b_aT_b$, where
\bea
X^b_a={1\over 1+ x}\left(
\begin{array}{ccc}
\left(1-x\varepsilon^2b^2\right) & x\varepsilon\left(a +\varepsilon bc\right) &
-x\varepsilon\left(\varepsilon a b -c\right) \\
-x\varepsilon\left(a -\varepsilon bc\right) & 1-x\varepsilon^2 c^2 &
x\varepsilon\left(b +\varepsilon ac\right) \\
-x\varepsilon\left(\varepsilon a b +c\right) & -x\varepsilon\left(b -\varepsilon ac\right) &
\left(1-x\varepsilon^2a^2\right)
\end{array}
\right)\,\,\,\,.
\eea
The matrix $Y_a^b$ is obtained from the expression of $X^b_a$ by replacing
$x$ with $-x$ and $\varepsilon$ with $-\varepsilon$.
\par
Finally, we should also mention that the integrability of the model (\ref{klimcik})
holds even when the Lie algebra ${\cal G}$ is non-semi-simple provided that
the bilinear form $<\,,\,>_{{\cal G}}$ is invertible
(the invertibility of the bilinear form is used in deriving the equations of motion). 
This is so because the proof of the integrability of the non-linear sigma model
(\ref{klimcik}) relies on the two relations (\ref{R-1}) and (\ref{R-2}), satisfied 
by the $R$ operator, and on the existence of an invertible bilinear form on 
the Lie algebra ${\cal{G}}$.
As an example, the non-semi-simple Lie algebra (\ref{NW-algebra}) with the
invertible bilinear form (\ref{NW-form}) possesses an $R$ operator of the form{\footnote{
There are other $R$ operators for the algebra (\ref{NW-algebra}). We have chosen 
to write down the simplest of them.}} 
\bea
R\vec{T}=\left(
\begin{array}{cccc}
0&1&0&-a\\
-1&0&0&-d \\
a&d&c&-b c \\
0&0&0&-c
\end{array} \right) 
\left(\begin{array}{c}
P_1\\
P_2\\
J\\
T
\end{array}\right)\,\,\,\,,
\eea
where $a$, $c$ and $d$ are arbitrary parameters ($b$ is the parameter already appearing in 
the invariant bilinear form (\ref{NW-form})). The resulting integrable non-linear 
sigma model, as written in (\ref{gen-pcm}), has a matrix $\Omega_{ab}$ given by
\bea
\Omega_{ab}={1\over 1+\varepsilon^2}\left(
\begin{array}{cccc}
1 & -\varepsilon & {\varepsilon \left(a-\varepsilon d\right)\over \left(1-\varepsilon c\right)} & 0 \\
\varepsilon & 1 & {\varepsilon \left(d+\varepsilon a\right)\over \left(1-\varepsilon c\right)} & 0 \\
-{\varepsilon\left(a+\varepsilon d\right) \over \left(1+\varepsilon c\right)} & 
{\varepsilon\left(\varepsilon a-d\right) \over\left(1+\varepsilon c\right)} &
{b+\varepsilon^2\left(b-d^2-a^2\right)\over\left(1-\varepsilon^2c^2\right)} &
{1+\varepsilon^2\over(1+\varepsilon c)} \\
0& 0& {1+\varepsilon^2\over \left(1-\varepsilon c\right)}& 0
\end{array}
\right)\,\,\,\,\,.
\label{my-NSS}
\eea
The Lax pair of this integrable non-linear sigma model is of the form
(\ref{gen-PCS-lax}) with $P_a=X^b_aT_b$ and $Q_a=Y^b_aT_b$ and where
\bea
X_a^b ={1\over 1+x}\left(
\begin{array}{cccc}
1 & x \varepsilon & 0 & {x \varepsilon(\varepsilon d-a) \over (1-\varepsilon c)}\\
-x \varepsilon & 1& 0 & {-x \varepsilon(d+\varepsilon a) \over (1-\varepsilon c)} \\
{x\varepsilon (a+\varepsilon d)\over (1+\varepsilon c)} &
{x\varepsilon (d-\varepsilon a)\over (1+\varepsilon c)} &
{1+\varepsilon c  +x\varepsilon(c-\varepsilon)\over (1+\varepsilon c)} &
-{x\varepsilon \left[b c(1+\varepsilon^2)-\varepsilon(a^2+d^2)\right] \over (1-\varepsilon^2 c^2)}\\
0 & 0& 0 & {1-\varepsilon c -x \varepsilon(c+\varepsilon)\over (1-\varepsilon c)}
\end{array}\right)
\eea
with the lower index counting the rows. The matrix $Y^b_a$ is obtained by substituting 
$x$ with $-x$ and $\varepsilon$ with $-\varepsilon$ in the expression of $X^b_a$.

\section{Duality and integrability}

In this section we will put forward a method 
for constructing new integrable non-linear sigma models
starting from already integrable ones. This is  
based on 
the concept of T-duality \cite{buscher}
that certain two-dimensional non-linear sigma have.  
\par
In order to illustrate the important role of T-duality in 
the construction of integrable non-linear sigma models,
we will start by considering a simple example. The discussion
of the general case will be treated somewhere else
\cite{mohammedi-in-pre} (Abelian T-duality in the context of integrability 
has been used in \cite{balog2} in a very particular non-linear sigma model).
We will consider the non-linear sigma model of the previous section
whose action is 
\be
S\left(g\right)=\int{\rm d}z{\rm d}\bar z \,
\Omega_{ab}\left(g^{-1}\dd g\right)^a
\left(g^{-1}\bar\dd g\right)^b
\,\,\,\,,
\label{original}
\ee
This theory has the symmetry transformation $g\longrightarrow Lg$,
where $L$ is a constant element of the Lie group to which $g$ belongs.
The duality transformations are found by first gauging this symmetry
and adding a Lagrange multiplier term which constrains the gauge
field strength to vanish \cite{nadual3,nadual4,amit}.
This results in the first order action
\bea
S_1\left(g,B,\bar B,\chi\right) &=&\int{\rm d}z{\rm d}\bar z \,
\left[\Omega_{ab}\left(g^{-1}\dd g+g^{-1}Bg\right)^a
\left(g^{-1}\bar\dd g+g^{-1}\bar Bg\right)^b\right.
\nonumber \\
&+& \chi_a\left(\dd \bar B^a -\bar\dd B^a + f^a_{bc}B^b\bar B^c\right)
\left.\right]
\,\,\,\,\,.
\eea
Here $B=B^aT_a$ and $\bar B=\bar B^aT_a$ are the gauge fields with the
gauge transformation $B\longrightarrow LBL^{-1}-\dd LL^{-1}$,
$\bar B\longrightarrow L\bar BL^{-1}-\bar\dd LL^{-1}$. The Lagrange 
multiplier $\chi_a$, transforming in the adjoint representation,  imposes the pure gauge condition 
$\dd \bar B^a -\bar\dd B^a + f^a_{bc}B^b\bar B^c=0$ whose solution
is $B=h^{-1}\dd h$ and $\bar B=h^{-1}\bar\dd h$. Upon replacing  this 
back into the action $S_1$ one gets the relation $S_1=S(hg)$ and by 
choosing $h=1$ (thanks to the local gauge symmetry $g\longrightarrow Lg$, $h\longrightarrow hL^{-1}$)
one concludes that the action $S_1$ is equivalent to the action $S$.
\par
The dual action is obtained by keeping the Lagrange multiplier and
eliminating, instead,  the gauge fields (through their equations of motion). This 
procedure yields, after the gauge choice $g=1$, the dual theory
\bea
\widetilde S\left(\chi\right) &=& \int{\rm d}z{\rm d}\bar z \,
\left(M^{-1}\right)^{ab}\dd\chi_a\bar\dd\chi_b
\nonumber \\
M_{ab} &\equiv&\Omega_{ab} +\chi_cf^c_{ab} \,\,\,\,.
\label{dual}
\eea
We will show now that if the original theory (\ref{original}) is integrable then 
its dual (\ref{dual}) is also integrable. It is convenient, for this purpose, to introduce
the two currents
\bea
J^a=\left(M^{-1}\right)^{ba}\dd \chi_b\,\,\,\,\,\,,\,\,\,\,\,\,
\bar J^a =-\left(M^{-1}\right)^{ab}\bar \dd \chi_b \,\,\,\,\,.
\eea
In terms of these, the equations of motion of the dual theory (\ref{dual}) are
\bea
\widetilde{\cal{E}}^a\equiv \dd \bar J^a - \bar \dd J^a +f^a_{bc}J^b\bar J^c=0\,\,\,\,.
\eea
We notice that these are the Bianchi identities 
(\ref{bianchi-original})
of the original theory with
$(J^a\,,\,\bar J^a)$ interchanged with $(A^a\,,\,\bar A^a)$. 
Furthermore, these currents satisfy the Bianchi identity 
(stemming from $\dd\bar\dd \chi_a -\bar\dd\dd\chi_a=0$)
\bea
\widetilde{\cal{B}}_c &\equiv&
 -{1\over 2}\left(\Omega_{cd}+\Omega_{dc}\right)
\left(\dd \bar J^d + \bar \dd J^d\right)
+\left[{1\over 2}\left(\Omega_{cd}-\Omega_{dc}\right)f^d_{ab}
+\left(\Omega_{ad}f^d_{bc} + \Omega_{db}f^d_{ac}\right)\right]J^a\bar J^b 
\nonumber \\
&-& \left[{1\over 2}\left(\Omega_{cd}-\Omega_{dc}\right)
+\chi_af^a_{cd}\right]\left(\dd \bar J^d - \bar \dd J^d
+ f^d_{be} J^b\bar J^e\right)
=0
\,\,\,\,,
\eea
Again these are a linear combination of the equations of motion 
(\ref{eom-original})
and the Bianchi identities (\ref{bianchi-original})
of the original sigma model 
with
$(J^a\,,\,\bar J^a)$ interchanged with $(A^a\,,\,\bar A^a)$.
\par
Therefore, for the dual non-linear sigma model, the linear combination
\bea
&\,& \widetilde{\cal{B}}_c 
+ \left[{1\over 2}\left(\Omega_{cd}-\Omega_{dc}\right)
+\chi_af^a_{cd}\right]\widetilde{\bf{\cal E}}^d
% =0=
 \nonumber \\
&=&  -{1\over 2}\left(\Omega_{cd}+\Omega_{dc}\right)
\left(\dd \bar J^d + \bar \dd J^d\right)
 +\left[{1\over 2}\left(\Omega_{cd}-\Omega_{dc}\right)f^d_{ab}
+\left(\Omega_{ad}f^d_{bc} + \Omega_{db}f^d_{ac}\right)\right]J^a\bar J^b 
\eea
takes exactly the form of the equations of motion (\ref{eom-original})
of the original theory with the exchange 
$(J^a\,,\,\bar J^a)$ $\longleftrightarrow $ $(A^a\,,\,\bar A^a)$.
\par
As already shown in equation (\ref{E-B}), the zero curvature condition 
stemming from a Lax pair is a linear combination of the equations of motion 
of the sigma model and some corresponding Bianchi identities.   
Moreover, the equations of motion and the Bianchi identities 
of the dual theory are simply linear combinations of those  
of the original theory with the exchange  
$(J^a\,,\,\bar J^a)$ $\longleftrightarrow $ $(A^a\,,\,\bar A^a)$. 
Consequently,  
if the original non-linear sigma model (\ref{original}) 
is integrable (namely, if equation (\ref{genPCS}) is satisfied) then
its dual theory (\ref{dual}) is also integrable. The Lax pair of the dual theory is 
\bea
&&\left(\dd + J^a P_a\right)\Psi=0
\nonumber\\
&&\left(\bar\dd + \bar J^b Q_b\right)\Psi=0
\,\,\,\,\,\,.
\label{dual-lax}
\eea
This is simply the Lax pair of the original theory with 
$(J^a\,,\,\bar J^a)$ and $(A^a\,,\,\bar A^a)$ interchanged. 
Indeed, if the original non-linear sigma model (\ref{original})
is integrable (that is when (\ref{genPCS}) is satisfied) then
the zero curvature condition of this last linear system is
\be
\widetilde{\cal{Z}} = R^a\widetilde{\cal{B}}_a
+\left[Q_a+\left(\Omega_{ba}+\chi_cf^c_{ba}\right)R^b\right]\widetilde{\cal{E}}^a\,\,\,\,,
\ee
where we have used the relation $P_a=Q_a+(\Omega_{ac}+\Omega_{ca})R^c$
and the integrability condition (\ref{genPCS}). We notice that the first term
vanishes identically and the second term yields the equations of motion 
of the dual non-linear sigma model. 

%The zero curvature condition of this linear system, upon using
%equation () and the Bianchi identities for the currents $J^a$ and $\bar J^a$, is
%\bea
%\left[\dd + J^a P_a\,,\,\bar\dd + \bar J^b Q_b\right]=
%\left(Q_d + M_{cd}R^c\right)\widetilde{\cal {E}}^d \,\,\,\,\,.
%\eea

\noindent
\vskip1.0cm
\noindent
{\it{\Large{Examples}}}
\vskip1.0cm

{\bf 1)} The first example which enters into the class 
of non-linear sigma models (\ref{original}) is of course the 
principal chiral non-linear sigma model for which
the matrix
$\Omega_{ab}=\eta_{ab}$, where $\eta_{ab}$ is the invariant bilinear
form of the underlying Lie algebra (that is, $\eta_{ab}f^b_{cd}+\eta_{cb}f^b_{ad}=0$).
Its corresponding dual non-linear sigma model
is written in (\ref{dual}) with
\be
M_{ab}=\eta_{ab}+\chi_c\,f^c_{ab}\,\,\,\,\,.
\label{dual-pcm}
\ee
The Lax pair of the dual of the principal chiral  non-linear sigma model is 
read from (\ref{lax1}), for $\kappa=0$, according to the above prescription. 
This is therefore given by
\bea
\left\{\dd + x\,\left[\left(M^{-1}\right)^{ab}\,\dd\chi_a\right]T_b\right\}\Psi &=&0
\nonumber\\
\left\{\bar\dd + {x\over 2x -1}
\,\left[-\left(M^{-1}\right)^{cd}\,\bar\dd\chi_d\right]T_c\right\}\Psi &=&0\,\,\,\,,
\eea
where $x$ is the spectral parameter.
\par
{\bf 2)} The second example we consider is the 
$SU(2)$-based non-linear sigma model 
whose corresponding matrix $\Omega_{ab}$ is given in (\ref{su2-omega})
and for which
\be
M_{ab}=\delta_{ab}\,L_b+\chi_c\,\epsilon_{cab}\,\,\,\,.
\ee
Its dual partner is given
by the action (\ref{dual}) with the matrix $M^{-1}$ 
explicitly given by
\bea
M^{-1}={1\over D}\left(
\begin{array}{lllll}
\chi_1^2+L_2L_3 & \, &\chi_1\chi_2-\chi_3 L_3 & \, &\chi_1\chi_3+\chi_2 L_2
\\
\chi_1\chi_2+\chi_3 L_3 & \, & \chi_2^2+L_1L_3 & \, & \chi_2\chi_3-\chi_1 L_1
\\
\chi_1\chi_3-\chi_2 L_2 & \, &\chi_2\chi_3+\chi_1 L_1 & \, & \chi_3^2+L_1L_2
\end{array} \right)\,\,\,\,,
\label{inv-M}
\eea
where
$D=L_1L_2L_3 +L_1\chi^2_1 +L_2\chi^2_2 +L_3\chi^2_3$.
The Lax pair of this dual theory is then given by
\bea
&&\left\{\dd + X^a_b \left[\left(M^{-1}\right)^{eb}\dd\chi_e\right] T_a\right\}\Psi=0
\nonumber\\
&&\left\{\bar\dd + Y^c_d \left[-\left(M^{-1}\right)^{df}\bar\dd\chi_f\right] T_c\right\}\Psi=0
\,\,\,\,\,\,,
\eea
where $T_a$ are the generators of the $SU(2)$ Lie algebra and the matrices $X^a_b$ and $Y^a_b$
are as given in (\ref{hlavaty}).

\section{Conclusion}

The question of classical integrability of two-dimensional non-linear sigma
models has been addressed in this paper. We have first focused on the issue
of representing the equations of motion of the non-linear sigma model
as a zero curvature condition of a linear system. This is regardless of whether
the Lax pair depended or not on a spectral parameter. This requirement
resulted in a master equation with some interesting geometrical properties.
In particular, it is shown that
in the case when the matrices involved in this equation are Lie algebra
valued matrices, this master equation is a generalisation of 
an equation encountered in the context of Poisson-Lie T-duality.
It is therefore hopeful that a general solution along the lines of 
\cite{klimcik1, klimcik2} might be found to this master equation.
\par
We have then put special emphasis on those constructions admitting 
a spectral parameter. Two situations emerged from this analyses. 
The first consists of those constructions where the spectral parameter
enters in a multiplicative way. The geometry of the integrable non-linear 
sigma models is in this case tractable. The isometry symmetry of these
sigma model plays an essential role and is responsible for their
integrability. Even in this simplified version, 
the general solution to the master equation remains a challenging
problem.  The second situation concerns those
constructions for which the spectral parameter does not manifest
itself in a multiplicative manner. Here, we have studied only the integrability of 
a generalisation of the principal chiral non-linear sigma model.
The master equation for these models, (\ref{genPCS}), 
has a certain Lie algebra structure and one hopes that this might 
be helpful in finding solutions. We have carried out a computer assisted 
study and found two new integrable non-linear sigma models. These are
given in (\ref{4-d}) and (\ref{5-d}) and are  based on two non-semi-simple 
Lie algebras of dimension four and five, respectively.
\par
Always in the context of the generalised principal chiral non-linear sigma model, 
we have given a brief summary of 
a new integrable non-linear sigma model that has recently been 
found by  Klim\v{c}\'{\i}k \cite{klimcik3}. This solution to the master
equation holds for any simple Lie algebra and relies on an $R$ operator 
which acts on the generators of the Lie algebra.
We have pointed out that the action of the $R$ operator as
given in  \cite{klimcik3}, equation (\ref{R-3}), is   
not the most general one. We have worked out the most general 
$R$ operator for the case of the $SU(2)$ Lie algebra. This has led 
to a more general integrable non-linear sigma model (\ref{my-su2}) 
in comparison with the one in (\ref{klimcik-su2}). However, 
the problem of  constructing the most general $R$ operator for other
Lie algebras is still an open issue. Furthermore, it is shown 
that the $R$ operator can also be extended to  non-semi-simple Lie algebras.
This was carried out for a particular example resulting in another
integrable non-linear sigma model (\ref{my-NSS}).     
  
\par
We have also shown that there is a connection between T-duality
and integrability of non-linear sigma models. More precisely,
if a non-linear sigma model is integrable and admits a T-duality
transformation then its dual is also integrable.
This might not sound surprising as two non-linear sigma models 
related by a T-duality transformation are by definition equivalent.
However, it is not obvious how to find the Lax pair
associated to the dual theory starting from the Lax pair
of the original non-linear sigma model. We have given 
here the recipe for this passage. T-duality is therefore a mean 
for constructing new integrable non-linear sigma models. 

\par
Among the open problems that could be addressed in the light of 
this work would be
the extension of the standard procedure of the dressing 
transformations \cite{zakharov,faddeev,babelon}
encountered in the principal chiral sigma model \cite{devchand, manas, spradlin}
to the new integrable models found in this paper. 
The construction of the conserved charges is also another challenging issue.

\par
Finally, we should mention that the study of the integrability
of non-linear sigma models carried out here  
could be of interest to string theory in its quest for integrable 
string backgrounds
\cite{tseytlin1,tseytlin2,gershun}.

\smallskip
\smallskip
\smallskip
\bigskip

\noindent $\underline{\hbox{\bf Acknowledgments}}$: 
I would like to thank P\'eter Forg\'acs, Max Niedermaier 
and Paul Sorba for very
useful discussions and Anastasia Doikou, Andreas Fring and Ctirad Klim\v{c}\'{\i}k for correspondence.
The pertinent remarks of an anonymous referee are also here acknowledged.

\end{document}